\RequirePackage{tikz}
\documentclass[sn-mathphys]{sn-jnl}

\usepackage{algorithm}
\usepackage{verbatim}
\usepackage{multirow}
\usepackage{subfig}
\usepackage{graphicx}
\usepackage{rotating}
\usepackage{csquotes}
\usepackage{physics}
\usepackage[title]{appendix}
\usepackage{url}
\usepackage[english]{babel}
\usepackage[latin1]{inputenc}
\usepackage{array}
\usepackage{multicol}
\usepackage{amsmath, dsfont}
\usepackage{program}
\usepackage{tikz}
\usetikzlibrary{shapes, positioning}
\usepackage{pgfplots}
\pgfplotsset{compat=1.16}

\jyear{2022}%

\theoremstyle{thmstyleone}%
\theoremstyle{thmstyletwo}%

\theoremstyle{thmstylethree}%
\newtheorem{definition}{Definition}%

\DeclareMathOperator*{\argmin}{arg\,min}

\begin{document}

\title[A heuristic the irregular strip packing problem]{A heuristic for solving the irregular strip packing problem with quantum optimization}

\author*{Paul-Amaury Matt$^1$\thanks{Corresponding author}}\email{paul-amaury.matt@ipa.fraunhofer.de}

\author{Marco Roth$^1$}\email{marco.roth@ipa.fraunhofer.de}

\affil{$^1$Department of Cyber Cognitive Intelligence (CCI), Fraunhofer Institute for Manufacturing Engineering and Automation IPA, Nobelstrasse 12, 70569 Stuttgart, Germany}

\abstract{We introduce a novel quantum computing heuristic for solving the irregular strip packing problem, a significant challenge in optimizing material usage across various industries. This problem involves arranging a set of irregular polygonal pieces within a fixed-height, rectangular container to minimize waste. Traditional methods heavily rely on manual optimization by specialists, highlighting the complexity and computational difficulty of achieving quasi-optimal layouts. The proposed algorithm employs a quantum-inspired heuristic that decomposes the strip packing problem into two sub-problems: ordering pieces via the traveling salesman problem and spatially arranging them in a rectangle packing problem. This strategy facilitates a novel application of quantum computing to industrial optimization, aiming to minimize waste and enhance material efficiency.
Experimental evaluations using both classical and quantum computational methods demonstrate the algorithm's efficacy. We evaluate the algorithm's performance using the quantum approximate optimization algorithm and the quantum alternating operator ansatz, through simulations and real quantum computers, and compare it to classical approaches.}


\maketitle

\section{Introduction}

The \emph{irregular strip packing problem} is a challenging and economically significant issue that involves fitting a set of polygonal pieces into a fixed-height, rectangular container in a manner that minimizes unused space, or waste. This task has broad implications, affecting industries ranging from fashion to automotive and electronics, where efficient material use is crucial for both cost reduction and environmental sustainability. Traditional approaches often rely on the expertise of specialists using CAD systems to achieve quasi-optimal layouts, highlighting the problem's complexity from both combinatorial and geometric perspectives. The quality of the placements produced by these specialist workers is high and according to \cite{gomes2006solving}, automatic solutions can only barely match this level of quality.

In this work, we introduce the \emph{Opus Incertum} algorithm, a quantum computing (QC) heuristic designed to efficiently tackle the irregular strip packing problem. While quantum optimization is a highly active field with applications across a wide range of problems \cite{abbas2023}, purely geometric problems such as the strip packing problem have received comparatively less attention in this domain. To make the problem tractable for QC, we decompose it into two sub-problems, a Traveling Salesman Problem (TSP) which shares similarities with the strip packing problem in terms of computational difficulty and a rectangular packing problem. The TSP can then be solved using QC algorithms such as the Quantum Approximate Optimization Algorithm \cite{farhi2000quantum} and a variation thereof called Quantum Alternating Operator Ansatz \cite{hadfield2019quantum}. Our method works towards applying QC to industrial optimization, offering a promising solution to the irregular strip packing problem by minimizing waste and optimizing material usage.

The remainder of the article is organised as follows. Section \ref{sec:related-work} is dedicated to related work. Section~\ref{sec:problem_formulation} formally introduces the mathematical formulation of the problem. In Section \ref{sec:heuristic-approach}, we introduce a QC-based heuristic. We then present in Section \ref{sec:experiments} the experimental results of our heuristic when executed on such computers with different methods available and compare the performance with existing classical methods. We finally draw our conclusions in Section \ref{sec:conclusion}.

\section{Related Work}
\label{sec:related-work}

A recent survey of mathematical models proposed in the last decades for nesting problems can be found in \cite{leao2020irregular}. The integer linear programming models of \cite{toledo2013dotted, rodrigues2017clique} use a grid for the discrete positioning of pieces. The mixed-integer linear programming models of \cite{scheithauer1993modeling, daniels1994multiple, dean2002minimizing, fischetti2009mixed, alvarez2013branch, cherri2016robust} assume continuous positioning of the pieces in the container. \cite{leao2016semi} introduces a mixed-integer linear programming model with semi-continuous positioning, i.e., continuous positioning on one axis and discrete positioning on the other. \cite{martinez2017matheuristics} is the first paper to tackle the problem of packing irregular shapes with unrestricted rotations. Other mathematical models that have been proposed are non-linear programming and constraint programming models. All these exact approaches which assume the pieces are polygons allow to find the optimal solution, but only work for small problems.

One of the main difficulties in solving the strip packing problem or other nesting problems is the necessity to  represent the problem and perform geometrical computations such as checking if pieces overlap or fit entirely inside the container. \cite{bennell2008geometry} provides a survey of the geometric tools available. These include pixel/ raster methods \cite{oliveira1993algorithms, segenreich1986optimal, babu2001generic}, direct trigonometry and the D-function \cite{preparata2012computational, konopasek1981mathematical, mahadevan1984optimization, ferreira1998flexible}, the no-fit polygon \cite{mahadevan1984optimization, milenkovic1991automatic, ghosh1991algebra, bennell2001hybridising, bennell2008geometry} and phi-functions \cite{stoyan2004phi, chernov2010mathematical}.

Heuristic methods have been designed in order to solve large instances of the irregular strip packing problem. An often used heuristic is the First Fit Decreasing Algorithm \cite{johnson1974worst}, which consists of packing the pieces into bins in the order of decreasing size. The bottom-left heuristic proposed by \cite{jakobs1996genetic} consists of  sequentially placing the pieces in as far as possible in the bottom-left corner without overlapping them with those previously positioned. An initial solution is usually obtained with the bottom-left heuristic and then improved via heuristics, meta-heuristics and compact and separation models. According to \cite{leao2020irregular}, some of the best results are obtained with the methods of \cite{sato2012algorithm, sato2019raster, elkeran2013new, leung2012extended, imamichi2009iterated, egeblad2007fast}.

\cite{layeb2012novel} was the first to propose a quantum inspired algorithm for solving the one-dimensional bin packing problem, where a set of items must be packed into a minimum number of bins. \cite{de2022hybrid, garcia2022comparative} proposed a quantum-classical hybrid approach to solve the one-dimensional bin packing problem using quantum annealing. \cite{terada2018ising} proposed an Ising model mapping to solve the two-dimensional regular packing problem, in which all pieces are assumed to have a rectangular shape. Using a quantum annealer, they were able to solve problem instances with up to eighteen rectangles.

\section{The irregular strip packing problem}
\label{sec:problem_formulation}

\begin{figure}[h]
\centering
\includegraphics[width=1.0\textwidth]{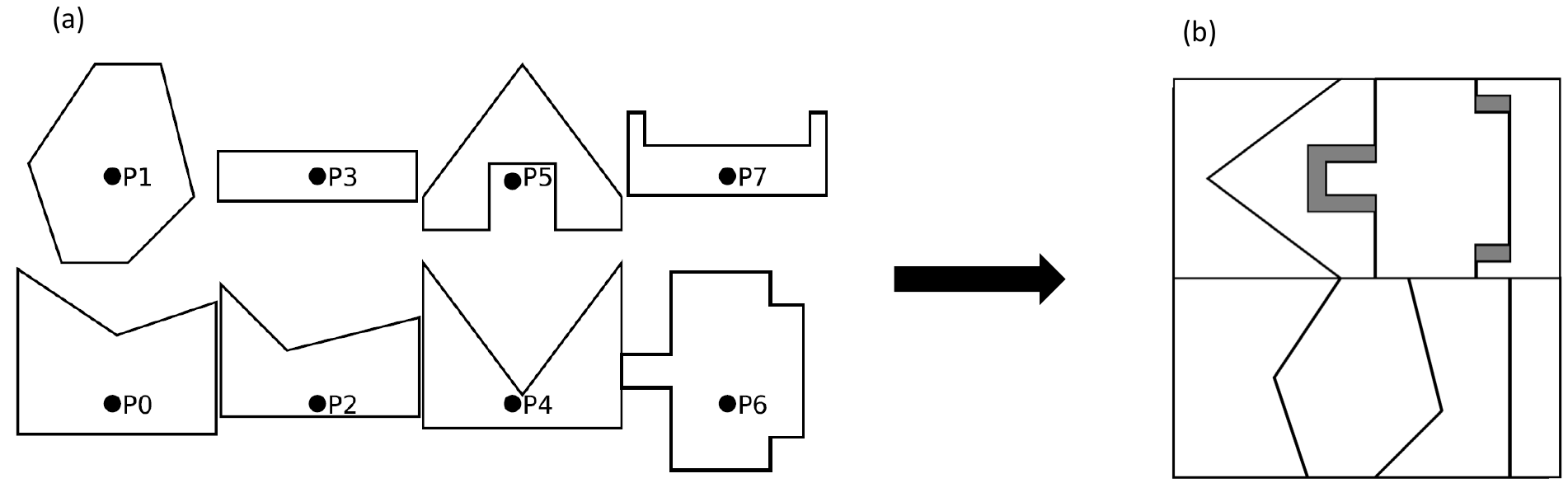}
\caption{(a) Set of eight pieces to pack. (b) Example of a hand-designed placement of the eight pieces.}
\label{fig:pieces_DEMO1}
\end{figure}

The irregular strip packing problem formally involves allocating a set of \(N\) pieces \(P_0, P_1, \ldots, P_{N-1}\) with polygonal shapes into a rectangular container \(C\) with a fixed height \(H\) and variable length \(L\). The pieces must be placed completely inside the container in such a way that they do not overlap. The pieces can be placed at any continuous location and can be freely oriented. The objective is to minimize the length $L$ of the container. Since all pieces must be placed inside the container and the height is fixed, minimizing the container length is equivalent to minimizing the unused area of the container, i.e., the surface that is unoccupied by the pieces. The unoccupied surface can be interpreted as \emph{waste} and the amount of waste can be quantified either by the area $W$ of that surface or by the ratio between $W$ and the total area $H \times L$ of the container. An example of a set with eight pieces is shown in Fig.~\ref{fig:pieces_DEMO1}(a). From a mathematical perspective, this problem as well as other irregular or regular packing problems, combine the combinatorial hardness of cutting and packing problems with the computational difficulty of enforcing the geometric non-overlap and containment constraints. A quasi-optimal layout for this set of eight pieces in a container with fixed height which has been designed by hand is provided in Fig.~\ref{fig:pieces_DEMO1}(b).

\section{The Opus Incertum Algorithm}
\label{sec:heuristic-approach}

Mathematical models proposed in the literature are solved monolithically, i.e., without use of decomposition methods thereby limiting the size of the problems that can be solved. \emph{Regular packing} (i.e. packing rectangular-shaped pieces) is an exception and constitutes a much simpler problem which can be solved efficiently. To make use of this, we propose a heuristic method where the irregular packing problem is reduced to a regular packing problem. We call the algorithm \emph{Opus Incertum Algorithm}\footnote{Opus Incertum, is a reference to an ancient Roman constructing technique of the same name that consists of irregularly shaped and randomly placed uncut stones inserted in a core of concrete. Vitruvius, in De architectura (Ten books on Architecture), favours opus incertum, deriding opus reticulatum (a similar technique using small pyramid-shaped instead of irregular stones) as more expensive and structurally of inferior quality.}. It consists of the following steps:
\begin{enumerate}
    \item Compute the geometrical compatibility between pieces
    \item Generate groups of geometrically compatible pieces
    \item Order the pieces in each group (corresponds to solving a TSP)
    \item Spatially arrange the pieces in each group into a compact rectangle
    \item Generate candidate partitions of the set of pieces
    \item Solve the rectangle packing problem for each partition
    \item Local optimization of the layout obtained for each partition
    \item Global optimization of the best layout
    \item Return the best layout
\end{enumerate}

\subsection{Definitions and details}
In the following we explain each of these steps in detail and introduce the necessary definitions. The final algorithm is summarized in Algorithm~\ref{alg:heuristic}.

\subsubsection{Geometrical compatibility between pieces}

Let us consider two pieces $P_i$ and $P_j$. Here, we assume that the position of $P_i$ is fixed and denote $(r, \theta, \phi)$ a placement with polar coordinates $(r, \theta)$ of the reference point of $P_j$ with respect to $P_i$ and $\phi$ the orientation of $P_j$. We want to find a placement $(r, \theta, \phi)$ of $P_j$ that avoids an overlap with $P_i$ while minimizing the waste. To make this statement mathematically precise, we define the waste as the surface that is inside the \emph{convex hull} of the set of vertices of the pieces. The optimal placement is then the one that avoids overlapping and yields a convex hull with minimum area. We introduce the concept of the \emph{no-fit function}. The no-fit function of two pieces $P_i$ and $P_j$ is the function $\mathrm{NFF}_{P_i, P_j}$ that takes as input a polar angle $\theta$ and returns the radius $r$ and orientation $\phi$ to optimally place $P_j$ with respect to $P_i$ given the rotation angle $\theta$. Once computed and stored, the no-fit function allows to find the placement that yields the convex hull with minimum area. The no-fit function is given by
\begin{equation} 
\mathrm{NFF}_{P_i, P_j}(\theta) = \argmin_{r \in \mathcal{R}, \phi \in \Phi} \left\lbrace \textrm{area}(\textrm{CH}[P_i, P_j(r,\theta,\phi)]) \, \vert \, P_i \cap P_j(r,\theta, \phi)=\emptyset \right\rbrace\,.\label{eq:nff}
\end{equation}
Here, $\mathrm{CH}(P_i, P_j)$ denotes the convex hull where the optimal convex hull $\mathrm{CH}^*(P_i, P_j)$ is given by  
\begin{equation}
    \mathrm{CH}^*(P_i, P_j) := \argmin_{\theta \in \Theta} \left\lbrace \textrm{area}(\textrm{CH}[P_i, P_j(r,\theta,\phi)]) \, \vert \, (r,\phi) = \mathrm{NFF}_{P_i, P_j}(\theta) \right\rbrace\,.
\end{equation}
Note that the absolute position of $P_i$ is irrelevant and the relative position of $P_j$ is implicitly given by the no-fit function. In the definitions above, $\Phi, \Theta$ are discrete sets of angles and $\mathcal{R}$ is a set of radii. For more information see Sec.~\ref{sec:clustering}.  

After optimally placing $P_j$ relative to $P_i$, the waste can be quantified by the difference between the area of the convex hull and the area of the two pieces. We will denote the waste as $d_{i,j}=\mathrm{d}(P_i, P_j)$, since we will later relate this quantity to the distance between two cities $i$ and $j$ in a TSP. With the definitions above, the waste  is given by
$$ d_{i,j} = \mathrm{d}(P_i, P_j) = \textrm{area}[\mathrm{CH}^*(P_i, P_j)] - \textrm{area}(P_i) - \textrm{area}(P_j)\,.$$

Furthermore, we define the following measure, which we will refer to as \emph{geometrical incompatibility} between $P_i$ and $P_j$. 
\begin{definition}[Distance and geometrical incompatibility/compatibility]
\begin{align}
    \mathrm{gi}(P_i, P_j) &= \frac{\mathrm{d}(P_i, P_j)}{\textrm{area}(\mathrm{CH}[P_i, P_j(r,\theta,\phi)])}\label{eq:geometrical_incompatibility} \\
    \mathrm{gc}(P_i, P_j) &= 1 - \mathrm{gi}(P_i, P_j)
\end{align}
\end{definition}

\begin{figure}[h]
    \centering
    \subfloat[]{\includegraphics[height=4.5cm]{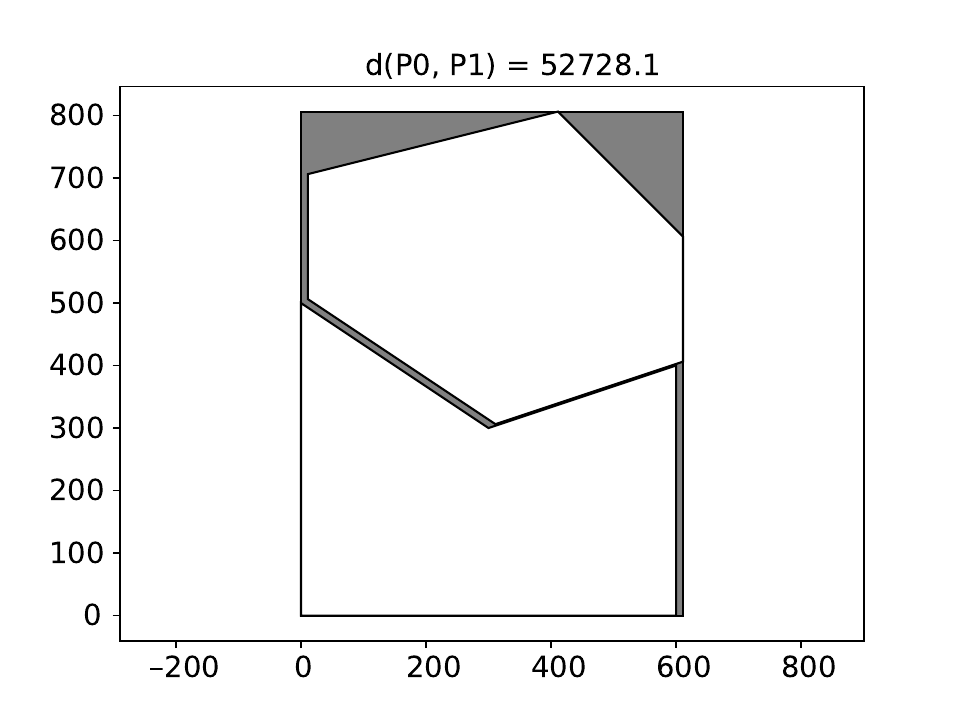}}
    \subfloat[]{\includegraphics[height=4.5cm]{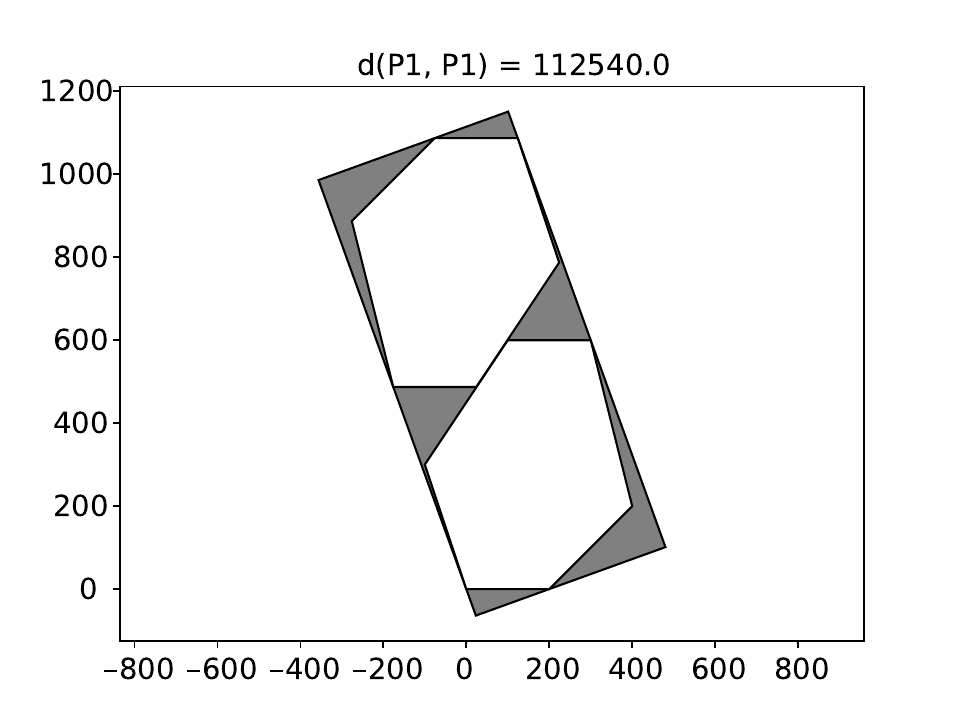}}

    \caption{The distance between two pieces and wasted area (grey). The axes and distances are given in arbitrary units.}
    \label{fig:incompatibility}
\end{figure}

Examples of distances are given in Fig.~\ref{fig:incompatibility}. The geometrical incompatibility is always bounded by $0$ and $1$. An incompatibility of $0$ means that the two pieces can be placed without waste. 

Our packing algorithm seeks to exploit the geometrical compatibility between pairs of pieces. When compatible pieces are placed together, the space between the pieces is small and this allows to minimize the wasted area in the container. The first step of the algorithm consists of computing the geometrical compatibility between all pairs of pieces, or equivalently, to compute the distance matrix, which we define as

\begin{definition}[Distance matrix]
The $N \times N$ square matrix $D$ with components $D_{i,j}=d_{i,j}$, where $d_{i,j}=\mathrm{d}(P_i, P_j)$ is the distance between pieces $P_i$ and $P_j$.
\end{definition}

\subsubsection{Generate groups of geometrically compatible pieces}

In the second step of the Opus Incertum algorithm, we generate groups of pieces in which the pieces are pairwise geometrically compatible. We then spatially arrange together the pieces that are geometrically compatible. For this, we use the \emph{single-linkage clustering} algorithm which is a hierarchical clustering method \cite{Everitt2011}. In the beginning, each element is in a cluster of its own. The agglomerative process consists of grouping the two clusters  that contain the closest pair of elements in each step. The clusters are then sequentially combined into larger clusters, until all elements are in the same cluster. We use the geometrical incompatibility measure  in Eq.~\eqref{eq:geometrical_incompatibility} to cluster the set of all pieces. The single-linkage clustering algorithm then produces clusters in which each piece has low geometrical incompatibility with at least one other pieces. The resulting set of clusters forms a partition of the set of all pieces. 

To this end, it is necessary to abort the agglomerative process before all the pieces belong to a single cluster. We achieve this in two ways. First, we introduce a threshold for the maximum linkage distance allowed for merging two clusters. To account for the arbitrariness of the threshold, we define several threshold values and obtain different partitions of the set of pieces. Second, we impose a maximum number of pieces per cluster, so that whenever this size is about to be exceeded, the agglomerative process terminates. 

\subsubsection{Ordering pieces in each cluster}

Consider a cluster $C_i$ and denote its pieces $P_{i_1}, \dotsc, P_{i_n}$. The goal is to place these pieces without overlap and as compactly as possible by exploiting their geometrical compatibility. Since the compatibility of pieces is only a binary measure, for $n > 2$ we do not know in general how large the area of the minimum bounding polygon for the $n$ cluster pieces is. Computing all possible placements that avoid overlap for $n$ pieces is not reasonable since it has an exponential scaling in terms of the number of pieces which becomes already problematic for small values of $n$. 

We therefore place the pieces one by one in the order of some sequence $P_{i_{\sigma(1)}}, \dotsc, P_{i_{\sigma(n)}}$. The number of possible permutations $\sigma$ for a set of $n$ elements is $n!$. To limit the computation, we choose the sequence with minimal total distance:
$$ \min_{\sigma} \sum_{i=1}^{n-1} \mathrm{d}(P_{i_{\sigma(i)}}, P_{i_{\sigma(i+1)}})$$
This sequence corresponds to the \emph{shortest Hamiltonian path}, defined as follows.

\begin{definition}[shortest Hamiltonian path]
Consider the fully connected, undirected and weighted graph with $n=\lvert V \rvert$ nodes $\{1, \dotsc, n \}$ and distances $d_{i,j}$ from node $i$ to $j$ as weights. A Hamiltonian path is a sequence $\sigma(1), \dotsc, \sigma(n)$ visiting all nodes of the graph exactly once, i.e., $\sigma$ is a permutation of the set $\{ 1, \dotsc, n\}$. The shortest Hamiltonian path is the Hamiltonian path with minimum total distance
$$ D(\sigma) = \sum_{i=1}^{n-1} d_{\sigma(i), \sigma(i+1)}$$
\end{definition}
 
Finding the shortest Hamiltonian path of a weighted graph is known as \emph{Traveling Salesman Problem} (TSP). When $n$ is small enough (roughly $n \leq 10$), the TSP is an NP-hard combinatorial optimization problem can be solved exactly and fast using brute force search. Solving larger instances quickly becomes computationally challenging. Quantum computing provides methods to solve this problem heuristically. In this work, we will focus on the Quantum Approximate Optimization Algorithm \cite{farhi2000quantum} and the a variation thereof called Quantum Alternating Operator Ansatz \cite{hadfield2019quantum}. The former requries the reformulation of the TSP into a quadratic unconstrained binary optimization (QUBO) problem. This is described in Appendix \ref{app:qubo}.

\subsubsection{Packing the pieces of each cluster}
\label{sec:clustering}

Once an order is obtained, the pieces in each cluster are packed as compactly as possible. For this we use a greedy algorithm that follows in the order of the chosen sequence $P_{i_{\sigma(1)}}, \dotsc, P_{i_{\sigma(n)}}$. We start by positioning the first piece $P_{i_{\sigma(1)}}$ with the default position $(0,0)$. For every subsequent step $k=2,\dotsc,n$, we then place the k-th piece $P_{i_{\sigma(k)}}$ by orbiting around the previous piece $P_{i_{\sigma(k-1)}}$ and searching for the position $(r,\theta)$ and orientation $\phi$ that minimizes the area of the bounding box of $P_{i_{\sigma(1)}}, \dotsc, P_{i_{\sigma(k)}}$ while avoiding overlap with the previously positioned pieces. 

\begin{figure}[ht]
	\centering
  \includegraphics[height=4.4cm]{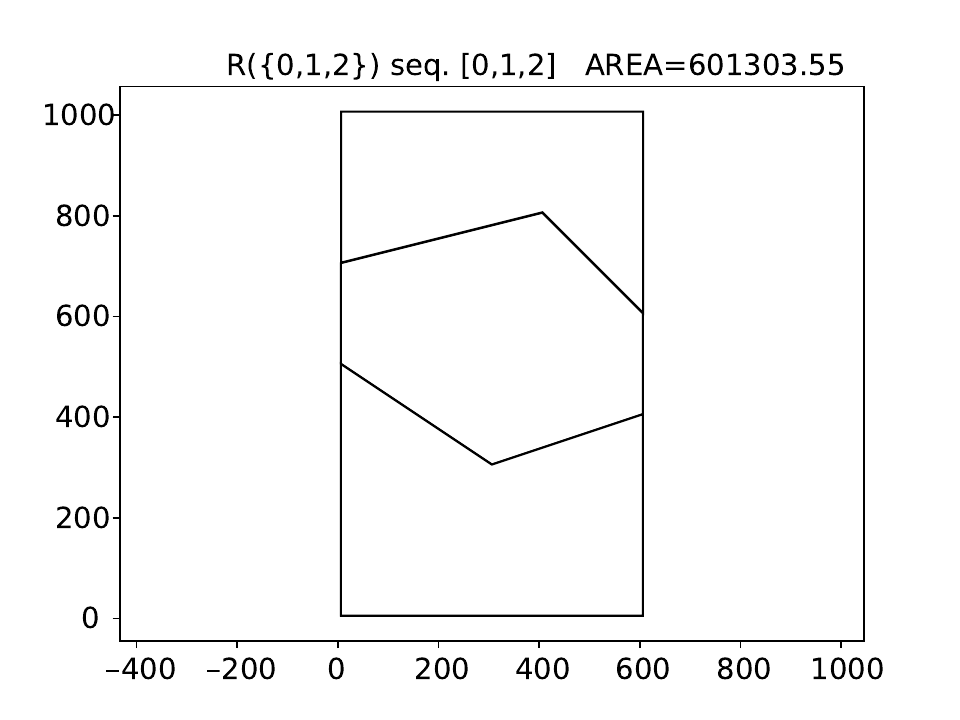}
  \includegraphics[height=4.4cm]{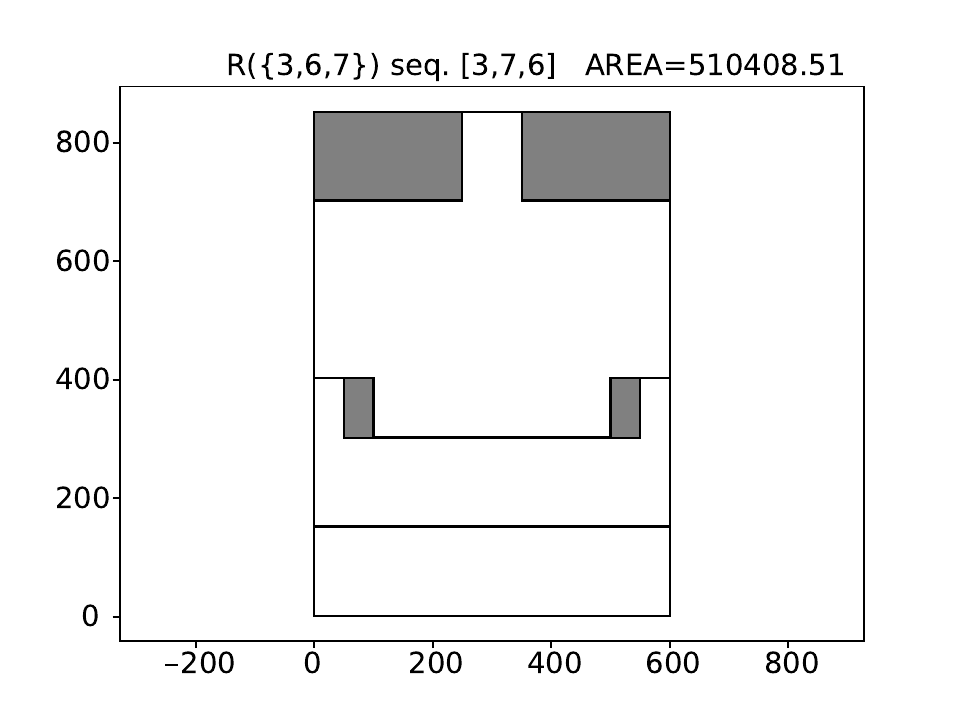}
  \includegraphics[height=4.4cm]{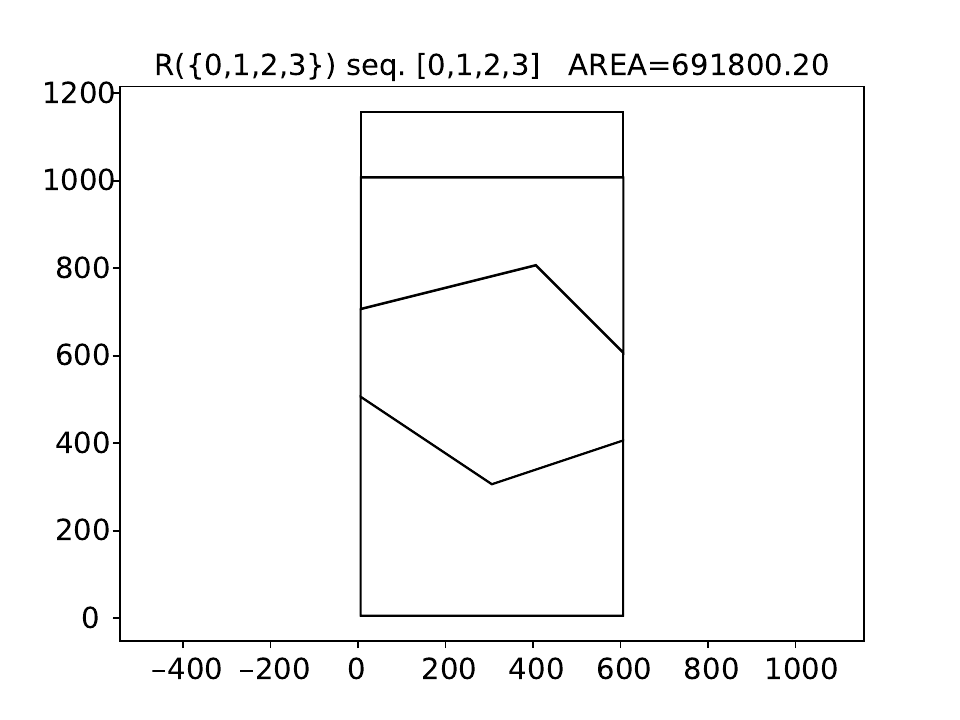}
  \includegraphics[height=4.4cm]{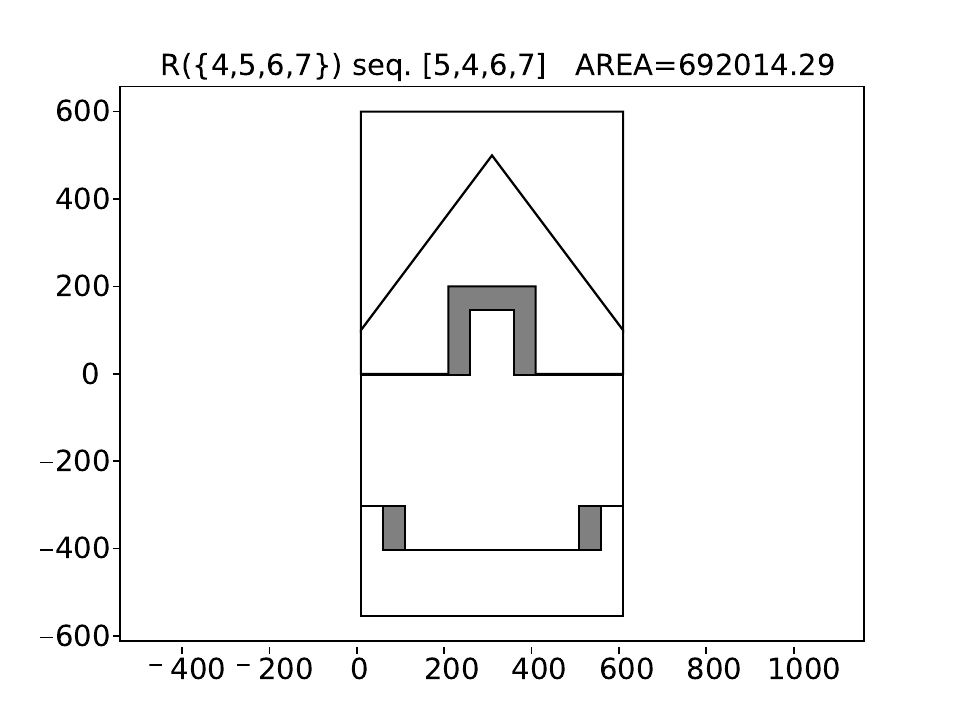}
  \caption{Packing of pieces in clusters. Axes are given in arbitrary units.}
	\label{fig:grouped_pieces}
\end{figure}

The term orbiting here means that we vary $\theta$ from a set of predefined angles $\Theta$. For each $\theta$, the no-fit function Eq.~\eqref{eq:nff} returns a radius $r$ and an angle $\phi$. If the $k$-th piece of the sequence can be placed relatively to the previous piece with position $(r,\theta,\phi)$ without overlap with the previously placed pieces, then we place the piece and proceed with the next piece. In the general case, however, the position returned by the no-fit function can lead to overlapping. We then increase the value of $r \in \mathcal{R}$ while keeping the orientation $\phi$ fixed, until the piece is far enough to avoid overlapping. We refer to this procedure as \emph{no-fit-function-based greedy packing}. Due to the no-fit function and the constraint of keeping the orientation fixed, the complexity of greedy packing is only $O([\vert \mathcal{R} \vert \cdot \vert \Theta \vert] \cdot [n-1])$. Once the pieces of a cluster are placed, we compute the corresponding bounding box, see Fig.~\ref{fig:grouped_pieces}.

\subsubsection{Partition filtering}

The number of partitions obtained may be large and lead to long computation times. We reduce the number of partitions used for packing the pieces by selecting the most promising ones. This can be done by computing a penalty for each partition and keeping only the best $n_{\textrm{partitions}}$ ones. A simple penalty score for some partition consists of the sum of areas of the rectangles. 

\subsubsection{Rectangle packing}

Each cluster is a group of pieces that has been placed together tightly into a bounding box of rectangular shape. In the rectangle packing step, these rectangles are allocated to the rectangular container with the help of an efficient rectangle packer. This is a well known problem \cite{Ibaraki2008} which formally consists of packing a set of $K$ rectangles $\mathcal{R}=\{ R_1, R_2, \dotsc, R_K\}$, where each rectangle $R_i$ has fixed length $l_i$ and height $h_i$ in the rectangular container $C$ of variable length $L$ and fixed height $H$. When all pieces are rectangular, the solution space for the irregular packing problem becomes finite \cite{bennell2008geometry}. If we denote $(x_i,y_i)$ the coordinates of the bottom left corner of rectangle $R_i$, the problem can be described as follows \cite{Ibaraki2008}:
\begin{align*}
    \mathrm{minimize} \quad & L \\
    \mathrm{subject} \, \mathrm{to} \quad & 0 \leq x_i \leq L - l_i, \quad 1 \leq i \leq K \\
    & 0 \leq y_i \leq H - h_i, \quad 1 \leq i \leq K \\
    & \mathrm{At} \, \mathrm{least} \,\mathrm{one} \,\mathrm{of} \, \mathrm{the} \, \mathrm{next} \,\mathrm{four} \, \mathrm{inequalities} \\
    & \mathrm{holds} \, \mathrm{for} \, \mathrm{every} \, \mathrm{pair} \, R_i \, \mathrm{and} \, R_j \, \mathrm{of} \, \mathrm{rectangles:} \\
    & x_i + l_i \leq x_j \\
    & x_j + l_j \leq x_i \\
    & y_i + h_i \leq y_j \\
    & y_j + h_j \leq y_i
\end{align*}

The first two constraints ensure that every rectangle is contained in the container. The remaining constraints express that rectanlges do not overlap. An illustration for the example in Fig.~\ref{fig:pieces_DEMO1} is given in Fig.~\ref{fig:layout_DEMO1}(a).

\subsubsection{Local optimization}

\begin{figure}
    \centering
    \includegraphics[width=5cm]{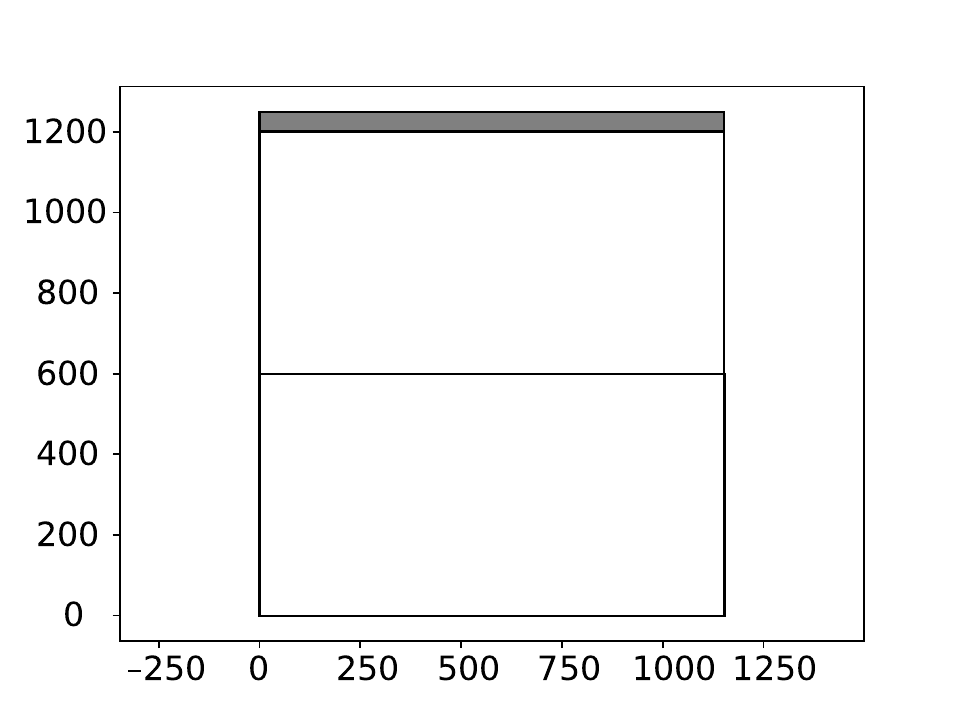}
    \includegraphics[width=5cm]{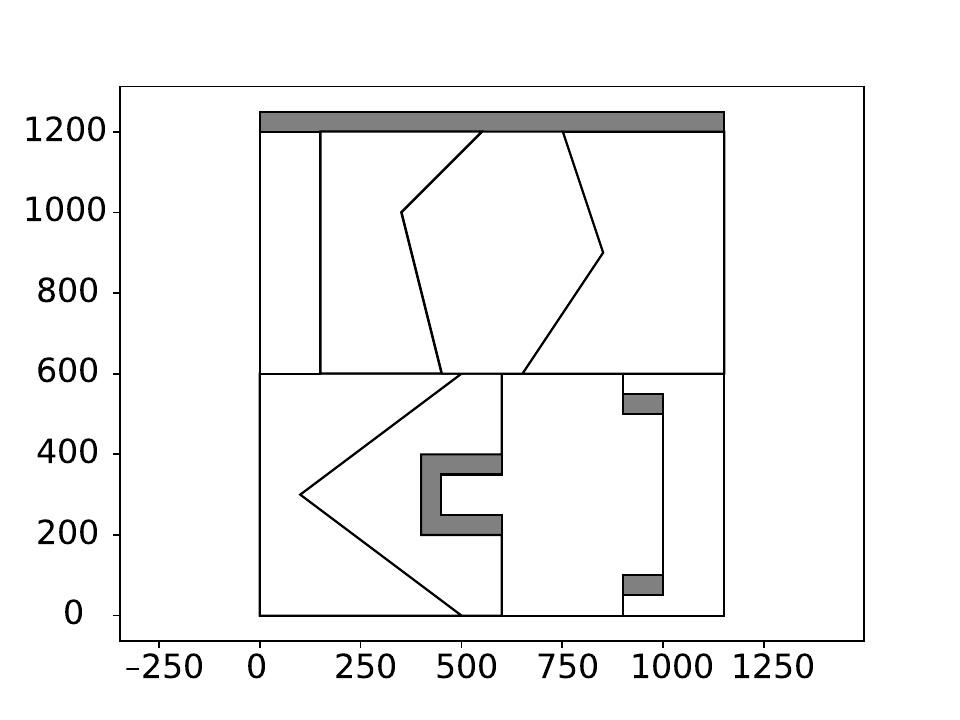}
    \caption{Results for the example shown in Fig.~\ref{fig:pieces_DEMO1}. (a) Rectangle packing for the . The rectangles packed are from left to right, bottom to top $R(\{0,1,2,3\})$, $R(\{4,5,6,7\})$.  (b) Layout obtained with Opus Incertum. The container has a length of $L=1151.89$. The percentage of waste is $6.59\%$. The axes are given in arbitrary units.}
    \label{fig:layout_DEMO1}
\end{figure}

The initial placement is obtained from positioning the rectangles in the container and substituting the rectangles for each cluster with the placement of the cluster pieces generated by greedy packing. Each such initial placement can be improved by local optimization. The objective of the local optimization procedure is to reduce the length needed for the container. In this procedure, the position of the pieces is variable, but their orientation is kept constant. The procedure consists of iterating through all the pieces from the bottom-left most to the top right most one and for each pieces, to apply translations of small amplitude (the granularity $\Delta r$ of $\mathcal{R}$) leftwards and/or downwards. Concretely, we apply translations with amplitude $\Delta r$ and direction given by one of the following four unit vectors:
$$ u_1 = \left[ \begin{array}{cc}
-1 \\
0 \end{array} \right] \quad u_2 = \frac{1}{\sqrt{2}} \left[ \begin{array}{cc}
-1 \\
-1 \end{array} \right] \quad u_3 = \frac{1}{\sqrt{2}} \left[ \begin{array}{cc}
-1 \\
+1 \end{array} \right] \quad u_4 = \left[ \begin{array}{cc}
0 \\
-1 \end{array} \right]$$
Translation are applied as long as the translated pieces does not overlap with the other pieces and remain completely inside the container. We cycle several times through all the pieces until no piece can be translated anymore. The result can be seen in Fig.~\ref{fig:layout_DEMO1}(b).

\subsubsection{Global optimization}

Local optimization reduces the gaps between neighbouring pieces, but it usually does not completely remove them. It is sometimes possible to make use of these gaps to fit the rightmost piece. By relocating the rightmost piece into one of these gaps, i.e., the piece which occupies a position the maximum value on the horizontal x-axis, we have a good chance to reduce the length needed for the container. This may not always be the case, as there may be several rightmost pieces. The relocation thus needs to be repeated until the currently selected rightmost piece cannot be relocated. To check if a piece can be relocated, we scan from left to right and bottom to top for a position and orientation of the piece to relocate and check if the piece can be inserted without exiting the boundaries of the container and without overlapping with one of the other pieces. Significant improvements of the container length may be obtained as a result of global optimization.

\subsection{The algorithm}

\begin{algorithm}[h]
\begin{algorithmic}[1]
\State \textbf{Inputs}: set of $N$ pieces $\mathcal{P}$, fixed container height $H$, maximum cluster size $n_{\max}$, number of partitions $n_{\textrm{partitions}}$
\State \textbf{Outputs}: optimal placement of the pieces and container length
\State compute the no-fit function $\mathrm{NFF}_{P_i,P_j}$ for every pair of pieces $P_i,P_j \in \mathcal{P}$
\State compute the distance and geometrical incompatibility matrices $D$ and $GI$
\State define the set of distance thresholds $\mathcal{T}(GI) \leftarrow \mathrm{values}(GI)$
\State compute a set of partitions of $\mathcal{P}$ by single-linkage clustering using the distance thresholds of $\mathcal{T}(GI)$ and maximum cluster sizes ranging from $1$ to $N$ pieces per cluster
\State initialize container length $L^* \leftarrow \infty$
\For{each partition $\mathcal{C}$}
\For{for each cluster $C \in \mathcal{C}$}
\State solve the TSP for the distance sub-matrix $D(C)$ 
\State let $s$ be the shortest Hamiltonian path obtained
\State pack greedily the pieces of $C$ in the order of $s$ 
\State pack greedily the pieces of $C$ in the reverse order of $s$
\State store the placement whose bounding box has the smallest area
\EndFor
\EndFor
\State keep from $\mathcal{C}$ the $n_{\textrm{partitions}}$ partitions with lowest penalty
\For{each partition $\mathcal{C}$}
\For{bins as relaxation of the rectangular container $H \times L^*$}
\State try to pack the selected bounding boxes in the bin
\If{a solution is found to the rectangle packing problem}
\State optimize the placement by local optimization
\State let $X$ be the optimized placement 
\State let $l$ and $h$ be the length and height needed by $X$
\If{$h \leq H$ and $l < L^*$}
\State update optimal placement $X^* \leftarrow X$ and length $L^* \leftarrow l$
\EndIf
\EndIf
\EndFor
\EndFor
\State optimize the placement $X^*$ and length $L^*$ by shifting the right-most piece to the bottom-left-most free area and repeat this step until no further improvement is possible
\State \Return optimal placement $X^*$ and length $L^*$
\end{algorithmic}
\caption{Opus Incertum}
\label{alg:heuristic}
\end{algorithm}

With the previous definitions, we can now state the formal algorithm which is shown in Algorithm~\ref{alg:heuristic}. Given the set of $N$ pieces $\mathcal{P}=\{ P_0, \dotsc, P_{N-1} \}$, we first compute and store the values of the no-fit functions Eq.~\eqref{eq:nff} for every pair of pieces $P_i$ and $P_j$. Note that the no-fit function does not need to be re-calculated if the pair of pieces is identical to a pair of pieces that has already been considered. This happens when pieces have identical shapes. Besides, after $\mathrm{NFF}{P_i,P_j}$ has been computed, we can derive directly values of the no-fit function $\mathrm{NFF}_{P_j,P_i}$ via 
\begin{equation}
\mathrm{NFF}_{P_j,P_i}(\theta) = (r, \phi) \text{ where } \mathrm{NFF}_{P_i,P_j}(\theta + 180 - \phi)=(r,-\phi)\,.
\end{equation}

Once all no-fit functions are determined, the $N \times N$ distance matrix $D$ can be quickly determined. Each distance $d_{i,j}$ is calculated as follows:
\begin{align*}
    \theta^* &\leftarrow \argmin_{\theta \in \Theta} \{ \mathrm{area}(\mathrm{CH}(P_i, P_j(r,\theta,\phi))) \, \vert \, (r,\phi) = \mathrm{NFF}_{P_i,P_j}(\theta) \} \\
    (r^*,\phi^*) &\leftarrow \mathrm{NFF}_{P_i,P_j}(\theta^*) \\
    d_{i,j} &\leftarrow  \mathrm{area}(\mathrm{CH}(P_i, P_j(r^*,\theta^*,\phi^*))) - \mathrm{area}(P_i) - \mathrm{area}(P_j(r^*,\theta^*,\phi^*)) \\
\end{align*}

Let $\mathcal{T}(GI)$ denote the set of threshold geometrical incompatibilities used for the single-linkage clustering algorithm. This set may be any subset of the set of coefficients of the distance matrix $GI$: 
$$\mathcal{T}(GI) \subseteq \mathrm{values}(GI) = \{ \mathrm{gi}_{i,j} \vert i \in \{0, \dotsc, N-1 \}, j \in \{0, \dotsc, N-1 \} \}\,.$$
In our algorithm definition, we choose $\mathcal{T}(GI) = \mathrm{values}(GI)$.

For every distance threshold $d_{\max}$ in $\mathcal{T}(GI)$ and maximum cluster size $n_{\max}$ from $1$ to $N$ (or some lower upper limit), partition the set of pieces using the single-linkage clustering algorithm and distance matrix $GI$. We obtain in this way a set of partitions of $\mathcal{P}$. We then initialize the variable $L^*$ which stores the container length for the best placement found. The initial value given is infinity.

For each partition $\mathcal{C}$, we consider every cluster $C$ in $\mathcal{C}$ and the distance of the pieces in $C$, noted $D(C)$, obtained directly as a sub-matrix of $D$. We then solve the TSP for $D(C)$ and get the shortest Hamiltonian path $s=i_1,\dotsc,i_n$ visiting once exactly all pieces of the cluster. Since the distance matrix $D$ and $D(C)$ are symmetric, the reverse path $s'=i_n, \dotsc, i_1$ has the same length and is another equally optimal solution of the TSP. We thus  greedily pack the pieces in each cluster $C$ either in the order of the sequence $s$ or in the order of $s'$. The placements and resulting bounding boxes differ in the general case. We select and store the placement that results in the bounding box with the smallest area of the two in memory.

For each partition $\mathcal{C}$, we try to pack the bounding boxes inside the container using a rectangle packer. More precisely, the packer gets the task of finding a feasible placement of all bounding boxes within a rectangular container called \emph{bin} whose dimensions coincide with the current best container found $H \times L^*$, where $L^*$ is the minimum length found so far. If the rectangle packer is successful, the bounding boxes are replaced by the pieces of the cluster they pack, resulting in a feasible placement of the set of all pieces $\mathcal{P}$ in a container of dimension $H \times L^*$. The placement can be further optimized using local optimization, resulting in a container length $L \leq L^*$. If $L$ is an improvement over $L^*$, then $L^*$ is updated with $L$ and we store the placement as the optimal placement $X^*$. If the length is not reduced, then no update is necessary. If the rectangle packer does not find a solution for this dimension, we relax the dimensions of the bin until a solution is found, by enlarging the length $L^*$ and/or the height $H$. Any solution found is optimized by local optimization and feasible solutions ($h \leq H$) are then compared to the optimal length $L^*$. Whenever the optimal length is improved, we update the variables $L^*$ and $X^*$.

Let $X^*$ be the best placement found after consideration of all the partitions. The length can be sometimes further optimized by global optimization, i.e., by iteratively relocating the right-most pieces to the left-most free gap in the container.

\section{Experimental Results}
\label{sec:experiments}

\begin{table}[t]
\caption{Performance of the Quantum Approximate Optimization Algorithm. The optimality is defined in Eq.~\eqref{eq:optimality}.}
    \centering
    \begin{tabular}{||c c c c||} 
 \hline
 Optimizer & $p$ & Optimality & Execution time \\ [0.5ex] 
 \hline\hline
COBYLA & 1 & 78.2\% & 2.5s \\
 & 2 & 76.6\% & 5.7s \\
 & 3 & 79.3\% & 9.0s \\
 & 4 & 77.1\% & 12.9s \\
 & 5 & 85.9\% & 17.3s \\
 \hline
BFGS & 1 & 68.7\% & 6.4s \\
 & 2 & 71.0\% & 16.8s \\
 & 3 & 73.6\% & 26.5s \\
 & 4 & 70.3\% & 44.5s \\
 & 5 & 73.1\% & 50.8s \\
 \hline 
L-BFGS-B & 1 & 68.7\% & 6.3s \\
 & 2 & 70.4\% & 17.5s \\
 & 3 & 73.6\% & 29.8s \\
 & 4 & 67.9\% & 38.2s \\
 & 5 & 73.1\% & 53.3s \\
\hline 
SLSQP & 1 & 70.9\% & 14.1s \\
 & 2 & 83.1\% & 46.5s \\
 & 3 & 72.1\% & 87.5s \\
 & 4 & 82.8\% & 142.2s \\
 & 5 & 84.3\% & 202.6s \\
\hline 
SPSA & 1 & 85.4\% & 23.4s \\
 & 2 & 81.6\% & 28.5s \\
 & 3 & 79.1\% & 36.5s \\
 & 4 & 74.0\% & 42.9s \\
 & 5 & 77.7\% & 51.4s \\
 \hline 
 \hline
\end{tabular}
\label{tab:tuning-approximate-optimization}
\end{table}
To evaluate the performance of our approach, we create two small problem instances which belong to the class of packing problems where we expect the Opus Incertum Algorithm to yield optimal results. The sets contain six and seven pieces, respectively. The pieces can be arranged in pairs or triples and form almost perfect rectangles when aligned. These can then be optimally packed inside the container by the rectangle packer. For reproducible, we provide the definition of the sets of pieces in Appendix \ref{app:puzzles}. We also evaluate our approach for five problem instances commonly used for benchmarking packing algorithms. The instances are referred to as SHAPES1, SHAPES2, SHIRTS, TROUSERS and SWIM and their definition can be found in \cite{oliveira2000topos}. Note however that these sets, which contain between 28 and 99 pieces, do not fulfil our requirements very well.

For each problem instance, we report the performance obtained with \texttt{SVGnest}\footnote{\url{https://github.com/Jack000/SVGnest}} compared to the results of Opus Incertum.  Our Python implementation, which is based on the Python modules Shapely\footnote{\url{https://pypi.org/project/shapely/}} and RectPack\footnote{\url{https://pypi.org/project/rectpack/}}.

\begin{table}[t]
\caption{Performance of the Quantum Alternating Operator Ansatz. The optimality is defined in Eq.~\eqref{eq:optimality}.}
    \centering
    \begin{tabular}{||c c c c||} 
 \hline
 Optimizer & $p$ & Optimality & Execution time \\ [0.5ex] 
 \hline\hline
COBYLA & 1 & 66.9\% & 6.8s \\
 & 2 & 70.9\% & 17.7s \\
 & 3 & 72.6\% & 33.0s \\
 & 4 & 84.6\% & 54.5s \\
 & 5 & 74.5\% & 79.0s \\
 \hline
BFGS & 1 & 61.6\% & 3.0s \\
 & 2 & 54.1\% & 6.4s \\
 & 3 & 61.3\% & 10.2s \\
 & 4 & 63.1\% & 14.4s \\
 & 5 & 63.5\% & 19.1s \\
 \hline 
L-BFGS-B & 1 & 62.9\% & 3.2s \\
 & 2 & 61.7\% & 6.6s \\
 & 3 & 61.3\% & 10.4s \\
 & 4 & 60.8\% & 14.4s \\
 & 5 & 62.9\% & 19.1s \\
\hline 
SLSQP & 1 & 64.2\% & 3.0s \\
 & 2 & 60.4\% & 6.4s \\
 & 3 & 61.8\% & 10.2s\\
 & 4 & 60.9\% & 14.4s \\
 & 5 & 66.1\% & 19.1s \\
\hline 
SPSA & 1 & 67.2\% & 38.2s \\
 & 2 & 76.1\% & 66.6s \\
 & 3 & 80.0\% & 96.4s \\
 & 4 & 72.9\% & 126.3s \\
 & 5 & 72.9\% & 157.3s \\
 \hline 
 \hline
\end{tabular}
\label{tab:tuning-alternating-operator}

\end{table}

To solve the TSPs, we implement one classical brute force search and two quantum algorithms: the Quantum Approximate Optimization Algorithm and the Quantum Alternating Operator Ansatz. The hyper-parameters for the quantum algorithms are the number of repetitions and the classical optimizer, which are set as follows. We randomly generate $30$ symmetric distance matrices of dimension $4 \times 4$, using coefficients uniformly distributed between $0$ and $1$. We run each algorithm on these problems using repetitions ranging from $1$ to $5$ and optimize the circuits using one of the following classical optimizers: COBYLA, BFGS, L-BFGS-B, SLSQP and SPSA. The quantum algorithms are then executed using Qiskit \cite{QiskitCommunity2017} on a quantum simulator without noise (qasm simulator). Each circuit is executed $1{\small,}000$ times and the result(s) with highest count is/are post-processed. Post-processing is necessary to ensure the validity of each solution. We call any Hamiltonian path, i.e., a sequence visiting each node exactly once a \emph{valid solution}. The post-processing is shown in Algorithm \ref{alg:post-processing}. In case multiple solutions are obtained, we compute the corresponding total distances and keep the Hamiltonian path $\sigma$ with smallest total distance $D(\sigma)$ as a unique solution. The performance for a path is measured by the optimality of its total distance, which we define as 
\begin{equation}
    1 - \frac{D(\sigma) - D_{\min}}{D_{\max} - D_{\min}}
    \label{eq:optimality}
\end{equation}
The optimality is averaged over the different TSPs.

The results are given in the Tables \ref{tab:tuning-approximate-optimization} and \ref{tab:tuning-alternating-operator}. It can be observed that in our experiments, the vanilla version of the Quantum Approximate Optmiziation Algorithm performs overall better while simulataneously requiring less execution time. As expected, the quality of the solution generally increases with an increasing number of iterations, although we observe some exceptions such as the SPSA variant of the Quantum Approximate Optmiziation Algorithm. For the following application of the Opus Incertum Algorithm, we proceed with a choice of the COBYLA optimizer and $p=5$ repetitions for the Quantum Approximate Optimization Algorithm, and the COBYLA optimizer and $p=4$ repetitions for the Quantum Alternating Operator Ansatz. 

\begin{algorithm}[t]
\begin{algorithmic}[1]
\State \textbf{Inputs}: a binary string $x_{0,0} \dotsc x_{i,p} \dotsc x_{n-1,n-1}$ of length $n^2$
\State \textbf{Outputs}: a hamiltonian path for the nodes $\{0, \dotsc, n-1\}$
\State for each node $i$, compute $\sum_{p=0}^{n-1} x_{i,p}$ and if the sum is strictly greater than $1$, select at random a step $p^*$ which $x_{i,p^*} = 1$ and set all the $x_{i,p}$ for $p \neq p^*$ to $0$
\State for each step $p$, compute $\sum_{i=0}^{n-1} x_{i,p}$ and if the sum is strictly greater than $1$, select at random a node $i^*$ which $x_{i,p^*} = 1$ and set all the $x_{i,p}$ for $i \neq i^*$ to $0$
\State for each node $i$, compute $\sum_{p=0}^{n-1} x_{i,p}$ and if the sum is null, select at random a step $p^*$ for which $\sum_j x_{j,p^*} = 0$ and set $x_{i,p^*}$ to $1$
\State for each step $p$, compute $\sum_{i=0}^{n-1} x_{i,p}$ and if the sum is null, select at random a node $i^*$ for which $\sum_{p'} x_{i^*,p'} = 0$ and set $x_{i^*,p}$ to $1$
\State let $\sigma=i_0, \dotsc, i_{n-1}$ be the path defined for any step $p$ by $i_p = i$ if $x_{i,p}=1$ 
\State \Return path $\sigma$
\end{algorithmic}
\caption{Post-processing of results from quantum circuits.}
\label{alg:post-processing}
\end{algorithm}

Once the hyper-parameters are set, we build variational circuits with $p$ repetitions and train the parameters of the QAOA algorithms with the chosen optimizer using a noiseless quantum simulator up to $n^2=16$ qubits. We compare these results to computations on the IBM quantum computer in Ehningen \cite{IBMQuantum}. As for the hyper-parameter optimization setting, we use $1{\small,}000$ shots per circuit, select the results with the highest count and post-process the bit-strings to obtain valid solutions.

\begin{figure}[t]
\tiny
    \centering
    \begin{tabular}{|p{3cm}||p{1.2cm}|p{1.2cm}|p{1.2cm}|}
         \hline
        \textbf{PROBLEM INSTANCE vs. METHOD} & \textbf{PUZZLE1} $6$ shapes, $6$ pieces, $H=750$ & \textbf{PUZZLE2} $7$ shapes, $7$ pieces, $H=420$ & \textbf{PUZZLE3} $12$ shapes, $12$ pieces, $H=1200$ \\
        \hline
        \hline
        Opus Incertum (Brute Force Search) & W=14.77\%, T=659.02s & W=6.55\%, T=927.17s & W=14.47\%, T=8784.24s\\
        \hline
        Opus Incertum (Quantum Approximate Optimization Algorithm, simulator) & W=14.77\%, T=645.50s & W=6.55\%, T=830.17s & W=18.76\%, T=8557.82s\\
        \hline
        Opus Incertum (Quantum Approximate Optimization Algorithm, quantum computer) & W=14.77\%, T=697.29s & W=6.70\%, T=899.36s & W=20.10\%, T=9155.27s \\
        \hline
        Opus Incertum (Quantum Alternating Operator Ansatz, simulator) & W=14.77\%, T=656.60s & W=6.55\%, T=902.98s & W=17.36\%, 9348.29s \\
        \hline
        Opus Incertum (Quantum Alternating Operator Ansatz, quantum computer) & W=14.77\%, T=705.43s & W=6.86\%, T=1069.09s & W=21.35\%, T=22877.37s \\
        \hline
        \hline
    \end{tabular}
    \caption{Performance results for PUZZLE1, PUZZLE2 and PUZZLE3. Results from real quantum computer are obtained from IBM Ehningen.}
    \label{fig:performance_puzzles}
\end{figure}

\begin{figure}[t]
\tiny
    \centering
    \begin{tabular}{|p{3cm}||p{1.2cm}|p{1.2cm}|p{1.2cm}|p{1.3cm}|p{1.2cm}|}
         \hline
        \textbf{PROBLEM INSTANCE vs. METHOD} & \textbf{SHAPES1} $4$ shapes, $43$ pieces, $H=400$ & \textbf{SHAPES2} $7$ shapes, $28$ pieces, $H=150$ & \textbf{SHIRTS} $8$ shapes, $99$ pieces, $H=400$ & \textbf{TROUSERS} $17$ shapes, $64$ pieces, $H=790$ & \textbf{SWIM} $10$ shapes, $48$ pieces, $H=5752$ \\
        \hline
        \hline
        Opus Incertum (Brute Force Search) & W=42.97\%, T=2137.24s & W=27.12\%, T=4981.99s & W=24.20\%, T=13279.78s & W=16.77\%, T=50639.89s & W=44.42\%, T=69377.16s \\
        \hline
        Opus Incertum (Quantum Approximate Optimization Algorithm, simulator) & W=42.97\%, T=3066.53s & W=27.12\%, T=5370.31s & W=22.33\%, T=15495.49s & W=16.77\%, T=39215.46s & W=49.78\%, T=32074.27s \\
        \hline
        Opus Incertum (Quantum Approximate Optimization Algorithm, quantum computer) & W=42.97\%, T=3424.93s & W=27.12\%, T=13373.23s & W=26.44\%, T=83871.25s & W=16.77\%, T=47787.09s & W=41.31\%, T=34753.92s \\
        \hline
        Opus Incertum (Quantum Alternating Operator Ansatz, simulator) & W=37.22\%, T=3356.84s & W=27.12\%, T=6611.17s & W=24.20\%, T=8294.80s & W=16.77\%, T=39171.81s & W=40.26\%, T=35808.40s \\
        \hline
        Opus Incertum (Quantum Alternating Operator Ansatz, quantum computer) & W=37.22\%, T=4462.86s & W=27.12\%, T=11551.44s & W=22.33\%, T=8930.77s & W=16.77\%, T=47459.27s & W=41.31\%, T=41310.82s \\
        \hline
        \hline
    \end{tabular}
    \caption{Performance results for SHAPES1, SHAPES2, SHIRTS, TROUSERS and SWIM. Results from real quantum computer are obtained from IBM Ehningen.}
    \label{fig:performance_benchmark}
\end{figure}

The performance results for PUZZLE1, PUZZLE2 and PUZZLE3 are shown in Fig.~\ref{fig:performance_puzzles} and for SHAPES1, SHAPES2, SHIRTS, TROUSERS and SWIM in Fig.~\ref{fig:performance_benchmark}. The results obtained with the Opus Incertum Algorithm depend on several factors, such as the number of rotation angles in $\Theta$ and orientation angles in $\Phi$, spatial granularity $\Delta r$, the spatial granularity of the grid used for relocating pieces, the maximum cluster size $n_{\max}$ allowed and the algorithm and hyper-parameter combination used to solve the TSP instances and the number of partitions. In our experiments, we allow  rotations in multiples of $5$ degrees which amounts to $72$ rotations in total. For PUZZLE1, PUZZLE2 and PUZZLE3, we restrict the rotations to multiples of $90$ degrees. For SHAPES1, SHAPES2, SHIRTS, SWIM and TROUSERS, we consider multiple of $45$ degrees, amounting to $8$ distinct rotations. The spatial granularity is set to $\Delta r = 5$. The grid is a uniform grid from the container of dimension $100 \times 100$. In our experimental setting, we limit the maximum cluster size to $4$ pieces as a higher number would then exceed the maximum number of $16$ qubits that a quantum simulator can handle. We set the number of partitions to $20$ for PUZZLE 1, $40$ for PUZZLE2 and $50$ for PUZZLE3. These parameters play a critical role in determining the algorithm's ability to find an optimal layout, with the choice of rotation angles directly affecting the potential for piece alignment and the granularity impacting the resolution of placement. The decision to use different parameters for different problem sets illustrates a tailored approach to optimization, trading off between computational demand and the quality of the solution.

Note, however, that better results than those displayed can be achieved by increasing for instance the number of rotations, increasing the maximum cluster size (up to $10$ pieces when solving the TSPs by brute force search) and the number of partitions, at the expense of longer computation times. The final placements for each problem instance can be found in Appendix~\ref{app:final_placements}.

It can be observed that solving the TSP with one of the quantum algorithms is often comparable to brute force solutions with no general tendency of a significant performance supremacy in either direction. Notably, the results obtained by solving the TSP on a real quantum computer are competitive with those obtained from a noiseless simulation, indicating a certain robustness in the Opus Incertum Algorithm. For example, in PUZZLE2, the waste percentage achieved by the Quantum Approximate Optimization Algorithm on a quantum computer ($W=6.70\%$, $T=899.36s$) closely mirrors that of the brute force search ($W=6.55\%$, $T=927.17s$), illustrating the quantum method's capacity to match classical performance levels. This parity is visible in other problem instances as well and suggests that while quantum computing offers a novel approach to problem-solving, its current stage of development shows comparable efficiency to classical methods for this specific application.

\section{Conclusion}
\label{sec:conclusion}

We have decomposed the NP-hard strip packing problem into two core problems, the TSP and the regular packing problem. In this work, we have solved the TSP classically and with quantum computing, using two different variants of the QAOA algorithm. Interestingly, the regular packing problem can be formulated in QUBO form and also be solved using quantum computing, as demonstrated recently in \cite{terada2018ising}. The practicality of the proposed algorithm as a quantum-classical hybrid or a purely classical, quantum-inspired method, is contingent on the advancements in quantum computing and the chosen approach for solving the underlying TSP. It is agnostic to the specific method employed for the TSP, meaning the performance significantly varies with the choice between novel quantum algorithms or classical heuristics. Furthermore, the selection of hyperparameters, particularly the granularity of angles and grids, introduces a crucial trade-off between the quality and performance of the solution. In comparing our algorithm's effectiveness, it is important to benchmark not only against brute force solutions but also against classical heuristics, which can offer a more efficient yet effective alternative. Another promising avenue for research is exploring scalability, particularly with regards to significantly increasing the number of pieces within each cluster. This is particularly interesting because, fundamentally, the Opus Incertum algorithm performs local optimization that becomes more global as the number of pieces within each cluster increases relative to the total number of clusters.

The proposed approach may be improved as follows. Other measures of geometrical compatibility may be developed and lead to more dense clusters of pieces. The algorithm complexity may be improved. Indeed, the number of partitions of the set of pieces used for generating candidate placements may be reduced, by considering only the most promising partitions. For instance, after computing all partitions, one could assign to each partition a loss simply defined as the total area of the boxes bounding the clustered pieces and choose the partitions with the smallest loss.

\section{Acknowledgements}

This work was part of the SEQUOIA project funded by the Minister for Economic Affairs, Labour and Tourism Baden-W\"{u}rttemberg. We also acknowledge use of the IBM Quantum Experience for this work. The views expressed are those of the authors and do not reflect the official policy or position of IBM or the IBM Quantum team.


\begin{thebibliography}{46}
\ifx \bisbn   \undefined \def \bisbn  #1{ISBN #1}\fi
\ifx \binits  \undefined \def \binits#1{#1}\fi
\ifx \bauthor  \undefined \def \bauthor#1{#1}\fi
\ifx \batitle  \undefined \def \batitle#1{#1}\fi
\ifx \bjtitle  \undefined \def \bjtitle#1{#1}\fi
\ifx \bvolume  \undefined \def \bvolume#1{\textbf{#1}}\fi
\ifx \byear  \undefined \def \byear#1{#1}\fi
\ifx \bissue  \undefined \def \bissue#1{#1}\fi
\ifx \bfpage  \undefined \def \bfpage#1{#1}\fi
\ifx \blpage  \undefined \def \blpage #1{#1}\fi
\ifx \burl  \undefined \def \burl#1{\textsf{#1}}\fi
\ifx \doiurl  \undefined \def \doiurl#1{\url{https://doi.org/#1}}\fi
\ifx \betal  \undefined \def \betal{\textit{et al.}}\fi
\ifx \binstitute  \undefined \def \binstitute#1{#1}\fi
\ifx \binstitutionaled  \undefined \def \binstitutionaled#1{#1}\fi
\ifx \bctitle  \undefined \def \bctitle#1{#1}\fi
\ifx \beditor  \undefined \def \beditor#1{#1}\fi
\ifx \bpublisher  \undefined \def \bpublisher#1{#1}\fi
\ifx \bbtitle  \undefined \def \bbtitle#1{#1}\fi
\ifx \bedition  \undefined \def \bedition#1{#1}\fi
\ifx \bseriesno  \undefined \def \bseriesno#1{#1}\fi
\ifx \blocation  \undefined \def \blocation#1{#1}\fi
\ifx \bsertitle  \undefined \def \bsertitle#1{#1}\fi
\ifx \bsnm \undefined \def \bsnm#1{#1}\fi
\ifx \bsuffix \undefined \def \bsuffix#1{#1}\fi
\ifx \bparticle \undefined \def \bparticle#1{#1}\fi
\ifx \barticle \undefined \def \barticle#1{#1}\fi
\bibcommenthead
\ifx \bconfdate \undefined \def \bconfdate #1{#1}\fi
\ifx \botherref \undefined \def \botherref #1{#1}\fi
\ifx \url \undefined \def \url#1{\textsf{#1}}\fi
\ifx \bchapter \undefined \def \bchapter#1{#1}\fi
\ifx \bbook \undefined \def \bbook#1{#1}\fi
\ifx \bcomment \undefined \def \bcomment#1{#1}\fi
\ifx \oauthor \undefined \def \oauthor#1{#1}\fi
\ifx \citeauthoryear \undefined \def \citeauthoryear#1{#1}\fi
\ifx \endbibitem  \undefined \def \endbibitem {}\fi
\ifx \bconflocation  \undefined \def \bconflocation#1{#1}\fi
\ifx \arxivurl  \undefined \def \arxivurl#1{\textsf{#1}}\fi
\csname PreBibitemsHook\endcsname

\bibitem{gomes2006solving}
\begin{barticle}
\bauthor{\bsnm{Gomes}, \binits{A.M.}},
\bauthor{\bsnm{Oliveira}, \binits{J.F.}}:
\batitle{Solving irregular strip packing problems by hybridising simulated annealing and linear programming}.
\bjtitle{European Journal of Operational Research}
\bvolume{171}(\bissue{3}),
\bfpage{811}--\blpage{829}
(\byear{2006})
\end{barticle}
\endbibitem

\bibitem{abbas2023}
\begin{botherref}
\oauthor{\bsnm{Abbas}, \binits{A.}},
\oauthor{\bsnm{Ambainis}, \binits{A.}},
\oauthor{\bsnm{Augustino}, \binits{B.}},
\oauthor{\bsnm{B{\"a}rtschi}, \binits{A.}},
\oauthor{\bsnm{Buhrman}, \binits{H.}},
\oauthor{\bsnm{Coffrin}, \binits{C.}},
\oauthor{\bsnm{Cortiana}, \binits{G.}},
\oauthor{\bsnm{Dunjko}, \binits{V.}},
\oauthor{\bsnm{Egger}, \binits{D.J.}},
\oauthor{\bsnm{Elmegreen}, \binits{B.G.}}, et al.:
Quantum optimization: Potential, challenges, and the path forward.
arXiv preprint arXiv:2312.02279
(2023)
\end{botherref}
\endbibitem

\bibitem{farhi2000quantum}
\begin{botherref}
\oauthor{\bsnm{Farhi}, \binits{E.}},
\oauthor{\bsnm{Goldstone}, \binits{J.}},
\oauthor{\bsnm{Gutmann}, \binits{S.}},
\oauthor{\bsnm{Sipser}, \binits{M.}}:
Quantum computation by adiabatic evolution.
arXiv preprint quant-ph/0001106
(2000)
\end{botherref}
\endbibitem

\bibitem{hadfield2019quantum}
\begin{barticle}
\bauthor{\bsnm{Hadfield}, \binits{S.}},
\bauthor{\bsnm{Wang}, \binits{Z.}},
\bauthor{\bsnm{O'gorman}, \binits{B.}},
\bauthor{\bsnm{Rieffel}, \binits{E.G.}},
\bauthor{\bsnm{Venturelli}, \binits{D.}},
\bauthor{\bsnm{Biswas}, \binits{R.}}:
\batitle{From the quantum approximate optimization algorithm to a quantum alternating operator ansatz}.
\bjtitle{Algorithms}
\bvolume{12}(\bissue{2}),
\bfpage{34}
(\byear{2019})
\end{barticle}
\endbibitem

\bibitem{leao2020irregular}
\begin{barticle}
\bauthor{\bsnm{Leao}, \binits{A.A.}},
\bauthor{\bsnm{Toledo}, \binits{F.M.}},
\bauthor{\bsnm{Oliveira}, \binits{J.F.}},
\bauthor{\bsnm{Carravilla}, \binits{M.A.}},
\bauthor{\bsnm{Alvarez-Vald{\'e}s}, \binits{R.}}:
\batitle{Irregular packing problems: A review of mathematical models}.
\bjtitle{European Journal of Operational Research}
\bvolume{282}(\bissue{3}),
\bfpage{803}--\blpage{822}
(\byear{2020})
\end{barticle}
\endbibitem

\bibitem{toledo2013dotted}
\begin{barticle}
\bauthor{\bsnm{Toledo}, \binits{F.M.}},
\bauthor{\bsnm{Carravilla}, \binits{M.A.}},
\bauthor{\bsnm{Ribeiro}, \binits{C.}},
\bauthor{\bsnm{Oliveira}, \binits{J.F.}},
\bauthor{\bsnm{Gomes}, \binits{A.M.}}:
\batitle{The dotted-board model: a new mip model for nesting irregular shapes}.
\bjtitle{International Journal of Production Economics}
\bvolume{145}(\bissue{2}),
\bfpage{478}--\blpage{487}
(\byear{2013})
\end{barticle}
\endbibitem

\bibitem{rodrigues2017clique}
\begin{barticle}
\bauthor{\bsnm{Rodrigues}, \binits{M.O.}},
\bauthor{\bsnm{Toledo}, \binits{F.M.}}:
\batitle{A clique covering mip model for the irregular strip packing problem}.
\bjtitle{Computers \& Operations Research}
\bvolume{87},
\bfpage{221}--\blpage{234}
(\byear{2017})
\end{barticle}
\endbibitem

\bibitem{scheithauer1993modeling}
\begin{barticle}
\bauthor{\bsnm{Scheithauer}, \binits{G.}},
\bauthor{\bsnm{Terno}, \binits{J.}}:
\batitle{Modeling of packing problems}.
\bjtitle{Optimization}
\bvolume{28}(\bissue{1}),
\bfpage{63}--\blpage{84}
(\byear{1993})
\end{barticle}
\endbibitem

\bibitem{daniels1994multiple}
\begin{botherref}
\oauthor{\bsnm{Daniels}, \binits{K.}},
\oauthor{\bsnm{Li}, \binits{Z.}},
\oauthor{\bsnm{Milenkovic}, \binits{V.}}:
Multiple containment methods
(1994)
\end{botherref}
\endbibitem

\bibitem{dean2002minimizing}
\begin{botherref}
\oauthor{\bsnm{Dean}, \binits{H.T.}}:
Minimizing waste in the 2-dimensional cutting stock problem
(2002)
\end{botherref}
\endbibitem

\bibitem{fischetti2009mixed}
\begin{barticle}
\bauthor{\bsnm{Fischetti}, \binits{M.}},
\bauthor{\bsnm{Luzzi}, \binits{I.}}:
\batitle{Mixed-integer programming models for nesting problems}.
\bjtitle{Journal of Heuristics}
\bvolume{15}(\bissue{3}),
\bfpage{201}--\blpage{226}
(\byear{2009})
\end{barticle}
\endbibitem

\bibitem{alvarez2013branch}
\begin{barticle}
\bauthor{\bsnm{Alvarez-Valdes}, \binits{R.}},
\bauthor{\bsnm{Martinez}, \binits{A.}},
\bauthor{\bsnm{Tamarit}, \binits{J.}}:
\batitle{A branch \& bound algorithm for cutting and packing irregularly shaped pieces}.
\bjtitle{International Journal of Production Economics}
\bvolume{145}(\bissue{2}),
\bfpage{463}--\blpage{477}
(\byear{2013})
\end{barticle}
\endbibitem

\bibitem{cherri2016robust}
\begin{barticle}
\bauthor{\bsnm{Cherri}, \binits{L.H.}},
\bauthor{\bsnm{Mundim}, \binits{L.R.}},
\bauthor{\bsnm{Andretta}, \binits{M.}},
\bauthor{\bsnm{Toledo}, \binits{F.M.}},
\bauthor{\bsnm{Oliveira}, \binits{J.F.}},
\bauthor{\bsnm{Carravilla}, \binits{M.A.}}:
\batitle{Robust mixed-integer linear programming models for the irregular strip packing problem}.
\bjtitle{European Journal of Operational Research}
\bvolume{253}(\bissue{3}),
\bfpage{570}--\blpage{583}
(\byear{2016})
\end{barticle}
\endbibitem

\bibitem{leao2016semi}
\begin{barticle}
\bauthor{\bsnm{Leao}, \binits{A.A.}},
\bauthor{\bsnm{Toledo}, \binits{F.M.}},
\bauthor{\bsnm{Oliveira}, \binits{J.F.}},
\bauthor{\bsnm{Carravilla}, \binits{M.A.}}:
\batitle{A semi-continuous mip model for the irregular strip packing problem}.
\bjtitle{International Journal of Production Research}
\bvolume{54}(\bissue{3}),
\bfpage{712}--\blpage{721}
(\byear{2016})
\end{barticle}
\endbibitem

\bibitem{martinez2017matheuristics}
\begin{barticle}
\bauthor{\bsnm{Martinez-Sykora}, \binits{A.}},
\bauthor{\bsnm{Alvarez-Vald{\'e}s}, \binits{R.}},
\bauthor{\bsnm{Bennell}, \binits{J.A.}},
\bauthor{\bsnm{Ruiz}, \binits{R.}},
\bauthor{\bsnm{Tamarit}, \binits{J.M.}}:
\batitle{Matheuristics for the irregular bin packing problem with free rotations}.
\bjtitle{European Journal of Operational Research}
\bvolume{258}(\bissue{2}),
\bfpage{440}--\blpage{455}
(\byear{2017})
\end{barticle}
\endbibitem

\bibitem{bennell2008geometry}
\begin{barticle}
\bauthor{\bsnm{Bennell}, \binits{J.A.}},
\bauthor{\bsnm{Oliveira}, \binits{J.F.}}:
\batitle{The geometry of nesting problems: A tutorial}.
\bjtitle{European journal of operational research}
\bvolume{184}(\bissue{2}),
\bfpage{397}--\blpage{415}
(\byear{2008})
\end{barticle}
\endbibitem

\bibitem{oliveira1993algorithms}
\begin{bbook}
\bauthor{\bsnm{Oliveira}, \binits{J.F.C.}},
\bauthor{\bsnm{Ferreira}, \binits{J.A.S.}}:
In: \beditor{\bsnm{Vidal}, \binits{R.V.V.}} (ed.)
\bbtitle{Algorithms for Nesting Problems},
pp. \bfpage{255}--\blpage{273}.
\bpublisher{Springer},
\blocation{Berlin, Heidelberg}
(\byear{1993})
\end{bbook}
\endbibitem

\bibitem{segenreich1986optimal}
\begin{barticle}
\bauthor{\bsnm{Segenreich}, \binits{S.A.}},
\bauthor{\bsnm{Braga}, \binits{L.M.P.F.}}:
\batitle{Optimal nesting of general plane figures: a monte carlo heuristical approach}.
\bjtitle{Computers \& Graphics}
\bvolume{10}(\bissue{3}),
\bfpage{229}--\blpage{237}
(\byear{1986})
\end{barticle}
\endbibitem

\bibitem{babu2001generic}
\begin{barticle}
\bauthor{\bsnm{Babu}, \binits{A.R.}},
\bauthor{\bsnm{Babu}, \binits{N.R.}}:
\batitle{A generic approach for nesting of 2-d parts in 2-d sheets using genetic and heuristic algorithms}.
\bjtitle{Computer-Aided Design}
\bvolume{33}(\bissue{12}),
\bfpage{879}--\blpage{891}
(\byear{2001})
\end{barticle}
\endbibitem

\bibitem{preparata2012computational}
\begin{botherref}
\oauthor{\bsnm{Preparata}, \binits{F.P.}},
\oauthor{\bsnm{Shamos}, \binits{M.I.}}:
Computational geometry: an introduction.
Springer
(2012)
\end{botherref}
\endbibitem

\bibitem{konopasek1981mathematical}
\begin{barticle}
\bauthor{\bsnm{Konopasek}, \binits{M.}}:
\batitle{Mathematical treatments of some apparel marking and cutting problems}.
\bjtitle{US Department of Commerce Report}
\bvolume{99}(\bissue{26}),
\bfpage{90857}--\blpage{10}
(\byear{1981})
\end{barticle}
\endbibitem

\bibitem{mahadevan1984optimization}
\begin{botherref}
\oauthor{\bsnm{Mahadevan}, \binits{A.}}:
Optimization in computer-aided pattern packing (marking, envelopes).
PhD thesis
(1984).
AAI8507009
\end{botherref}
\endbibitem

\bibitem{ferreira1998flexible}
\begin{botherref}
\oauthor{\bsnm{Ferreira}, \binits{J.}},
\oauthor{\bsnm{Alves}, \binits{J.}},
\oauthor{\bsnm{Albuquerque}, \binits{C.}},
\oauthor{\bsnm{Oliveira}, \binits{J.}},
\oauthor{\bsnm{Ferreira}, \binits{J.}},
\oauthor{\bsnm{Matos}, \binits{J.}}:
A flexible custom computing machine for nesting problems.
Proceedings of XIII DCIS, Madrid, Spain,
348--354
(1998)
\end{botherref}
\endbibitem

\bibitem{milenkovic1991automatic}
\begin{bchapter}
\bauthor{\bsnm{Milenkovic}, \binits{V.}},
\bauthor{\bsnm{Daniels}, \binits{K.}},
\bauthor{\bsnm{Li}, \binits{Z.}}:
\bctitle{Automatic marker making}.
In: \bbtitle{Proceedings of the Third Canadian Conference on Computational Geometry},
pp. \bfpage{243}--\blpage{246}
(\byear{1991}).
\bcomment{Simon Fraser University}
\end{bchapter}
\endbibitem

\bibitem{ghosh1991algebra}
\begin{barticle}
\bauthor{\bsnm{Ghosh}, \binits{P.K.}}:
\batitle{An algebra of polygons through the notion of negative shapes}.
\bjtitle{CVGIP: Image Understanding}
\bvolume{54}(\bissue{1}),
\bfpage{119}--\blpage{144}
(\byear{1991})
\end{barticle}
\endbibitem

\bibitem{bennell2001hybridising}
\begin{barticle}
\bauthor{\bsnm{Bennell}, \binits{J.A.}},
\bauthor{\bsnm{Dowsland}, \binits{K.A.}}:
\batitle{Hybridising tabu search with optimisation techniques for irregular stock cutting}.
\bjtitle{Management Science}
\bvolume{47}(\bissue{8}),
\bfpage{1160}--\blpage{1172}
(\byear{2001})
\end{barticle}
\endbibitem

\bibitem{stoyan2004phi}
\begin{barticle}
\bauthor{\bsnm{Stoyan}, \binits{Y.}},
\bauthor{\bsnm{Scheithauer}, \binits{G.}},
\bauthor{\bsnm{Gil}, \binits{N.}},
\bauthor{\bsnm{Romanova}, \binits{T.}}:
\batitle{Phi-functions for complex 2d-objects}.
\bjtitle{Quarterly Journal of the Belgian, French and Italian Operations Research Societies}
\bvolume{2}(\bissue{1}),
\bfpage{69}--\blpage{84}
(\byear{2004})
\end{barticle}
\endbibitem

\bibitem{chernov2010mathematical}
\begin{barticle}
\bauthor{\bsnm{Chernov}, \binits{N.}},
\bauthor{\bsnm{Stoyan}, \binits{Y.}},
\bauthor{\bsnm{Romanova}, \binits{T.}}:
\batitle{Mathematical model and efficient algorithms for object packing problem}.
\bjtitle{Computational Geometry}
\bvolume{43}(\bissue{5}),
\bfpage{535}--\blpage{553}
(\byear{2010})
\end{barticle}
\endbibitem

\bibitem{johnson1974worst}
\begin{barticle}
\bauthor{\bsnm{Johnson}, \binits{D.S.}},
\bauthor{\bsnm{Demers}, \binits{A.}},
\bauthor{\bsnm{Ullman}, \binits{J.D.}},
\bauthor{\bsnm{Garey}, \binits{M.R.}},
\bauthor{\bsnm{Graham}, \binits{R.L.}}:
\batitle{Worst-case performance bounds for simple one-dimensional packing algorithms}.
\bjtitle{SIAM Journal on computing}
\bvolume{3}(\bissue{4}),
\bfpage{299}--\blpage{325}
(\byear{1974})
\end{barticle}
\endbibitem

\bibitem{jakobs1996genetic}
\begin{barticle}
\bauthor{\bsnm{Jakobs}, \binits{S.}}:
\batitle{On genetic algorithms for the packing of polygons}.
\bjtitle{European journal of operational research}
\bvolume{88}(\bissue{1}),
\bfpage{165}--\blpage{181}
(\byear{1996})
\end{barticle}
\endbibitem

\bibitem{sato2012algorithm}
\begin{barticle}
\bauthor{\bsnm{Sato}, \binits{A.K.}},
\bauthor{\bsnm{Martins}, \binits{T.C.}},
\bauthor{\bsnm{Tsuzuki}, \binits{M.S.G.}}:
\batitle{An algorithm for the strip packing problem using collision free region and exact fitting placement}.
\bjtitle{Computer-Aided Design}
\bvolume{44}(\bissue{8}),
\bfpage{766}--\blpage{777}
(\byear{2012})
\end{barticle}
\endbibitem

\bibitem{sato2019raster}
\begin{barticle}
\bauthor{\bsnm{Sato}, \binits{A.K.}},
\bauthor{\bsnm{Martins}, \binits{T.C.}},
\bauthor{\bsnm{Gomes}, \binits{A.M.}},
\bauthor{\bsnm{Tsuzuki}, \binits{M.S.G.}}:
\batitle{Raster penetration map applied to the irregular packing problem}.
\bjtitle{European Journal of Operational Research}
\bvolume{279}(\bissue{2}),
\bfpage{657}--\blpage{671}
(\byear{2019})
\end{barticle}
\endbibitem

\bibitem{elkeran2013new}
\begin{barticle}
\bauthor{\bsnm{Elkeran}, \binits{A.}}:
\batitle{A new approach for sheet nesting problem using guided cuckoo search and pairwise clustering}.
\bjtitle{European Journal of Operational Research}
\bvolume{231}(\bissue{3}),
\bfpage{757}--\blpage{769}
(\byear{2013})
\end{barticle}
\endbibitem

\bibitem{leung2012extended}
\begin{barticle}
\bauthor{\bsnm{Leung}, \binits{S.C.}},
\bauthor{\bsnm{Lin}, \binits{Y.}},
\bauthor{\bsnm{Zhang}, \binits{D.}}:
\batitle{Extended local search algorithm based on nonlinear programming for two-dimensional irregular strip packing problem}.
\bjtitle{Computers \& Operations Research}
\bvolume{39}(\bissue{3}),
\bfpage{678}--\blpage{686}
(\byear{2012})
\end{barticle}
\endbibitem

\bibitem{imamichi2009iterated}
\begin{barticle}
\bauthor{\bsnm{Imamichi}, \binits{T.}},
\bauthor{\bsnm{Yagiura}, \binits{M.}},
\bauthor{\bsnm{Nagamochi}, \binits{H.}}:
\batitle{An iterated local search algorithm based on nonlinear programming for the irregular strip packing problem}.
\bjtitle{Discrete Optimization}
\bvolume{6}(\bissue{4}),
\bfpage{345}--\blpage{361}
(\byear{2009})
\end{barticle}
\endbibitem

\bibitem{egeblad2007fast}
\begin{barticle}
\bauthor{\bsnm{Egeblad}, \binits{J.}},
\bauthor{\bsnm{Nielsen}, \binits{B.K.}},
\bauthor{\bsnm{Odgaard}, \binits{A.}}:
\batitle{Fast neighborhood search for two-and three-dimensional nesting problems}.
\bjtitle{European Journal of Operational Research}
\bvolume{183}(\bissue{3}),
\bfpage{1249}--\blpage{1266}
(\byear{2007})
\end{barticle}
\endbibitem

\bibitem{layeb2012novel}
\begin{barticle}
\bauthor{\bsnm{Layeb}, \binits{A.}},
\bauthor{\bsnm{Boussalia}, \binits{S.R.}}:
\batitle{A novel quantum inspired cuckoo search algorithm for bin packing problem}.
\bjtitle{International Journal of Information Technology and Computer Science}
\bvolume{4}(\bissue{5}),
\bfpage{58}--\blpage{67}
(\byear{2012})
\end{barticle}
\endbibitem

\bibitem{de2022hybrid}
\begin{botherref}
\oauthor{\bparticle{de} \bsnm{Andoin}, \binits{M.G.}},
\oauthor{\bsnm{Osaba}, \binits{E.}},
\oauthor{\bsnm{Oregi}, \binits{I.}},
\oauthor{\bsnm{Villar-Rodriguez}, \binits{E.}},
\oauthor{\bsnm{Sanz}, \binits{M.}}:
Hybrid quantum-classical heuristic for the bin packing problem.
arXiv preprint arXiv:2204.05637
(2022)
\end{botherref}
\endbibitem

\bibitem{garcia2022comparative}
\begin{botherref}
\oauthor{\bsnm{Garcia-de-Andoin}, \binits{M.}},
\oauthor{\bsnm{Oregi}, \binits{I.}},
\oauthor{\bsnm{Villar-Rodriguez}, \binits{E.}},
\oauthor{\bsnm{Osaba}, \binits{E.}},
\oauthor{\bsnm{Sanz}, \binits{M.}}:
Comparative benchmark of a quantum algorithm for the bin packing problem.
arXiv preprint arXiv:2207.07460
(2022)
\end{botherref}
\endbibitem

\bibitem{terada2018ising}
\begin{bchapter}
\bauthor{\bsnm{Terada}, \binits{K.}},
\bauthor{\bsnm{Oku}, \binits{D.}},
\bauthor{\bsnm{Kanamaru}, \binits{S.}},
\bauthor{\bsnm{Tanaka}, \binits{S.}},
\bauthor{\bsnm{Hayashi}, \binits{M.}},
\bauthor{\bsnm{Yamaoka}, \binits{M.}},
\bauthor{\bsnm{Yanagisawa}, \binits{M.}},
\bauthor{\bsnm{Togawa}, \binits{N.}}:
\bctitle{An ising model mapping to solve rectangle packing problem}.
In: \bbtitle{2018 International Symposium on VLSI Design, Automation and Test (VLSI-DAT)},
pp. \bfpage{1}--\blpage{4}
(\byear{2018}).
\bcomment{IEEE}
\end{bchapter}
\endbibitem

\bibitem{Everitt2011}
\begin{bbook}
\bauthor{\bsnm{Everitt}, \binits{B.S.}},
\bauthor{\bsnm{Landau}, \binits{S.}},
\bauthor{\bsnm{Leese}, \binits{M.}},
\bauthor{\bsnm{Stahl}, \binits{D.}}:
\bbtitle{Cluster Analysis},
pp. \bfpage{73}--\blpage{75}.
\bpublisher{John Wiley {\&} Sons, Ltd},
\blocation{Hoboken}
(\byear{2011})
\end{bbook}
\endbibitem

\bibitem{Ibaraki2008}
\begin{bbook}
\bauthor{\bsnm{Ibaraki}, \binits{T.}},
\bauthor{\bsnm{Imahori}, \binits{S.}},
\bauthor{\bsnm{Yagiura}, \binits{M.}}:
In: \beditor{\bsnm{Blum}, \binits{C.}},
\beditor{\bsnm{Aguilera}, \binits{M.J.B.}},
\beditor{\bsnm{Roli}, \binits{A.}},
\beditor{\bsnm{Sampels}, \binits{M.}} (eds.)
\bbtitle{Hybrid Metaheuristics for Packing Problems},
pp. \bfpage{185}--\blpage{219}.
\bpublisher{Springer},
\blocation{Berlin, Heidelberg}
(\byear{2008})
\end{bbook}
\endbibitem

\bibitem{oliveira2000topos}
\begin{barticle}
\bauthor{\bsnm{Oliveira}, \binits{J.F.}},
\bauthor{\bsnm{Gomes}, \binits{A.M.}},
\bauthor{\bsnm{Ferreira}, \binits{J.S.}}:
\batitle{Topos--a new constructive algorithm for nesting problems}.
\bjtitle{OR-Spektrum}
\bvolume{22}(\bissue{2}),
\bfpage{263}--\blpage{284}
(\byear{2000})
\end{barticle}
\endbibitem

\bibitem{QiskitCommunity2017}
\begin{botherref}
\oauthor{\bsnm{{Qiskit Community}}}:
Qiskit: {{An}} Open-Source Framework for Quantum Computing
(2017).
\doiurl{10.5281/zenodo.2562110}.
\url{https://github.com/Qiskit/qiskit}
\end{botherref}
\endbibitem

\bibitem{IBMQuantum}
\begin{botherref}
\oauthor{\bsnm{{IBM Quantum}}}
\url{https://quantum-computing.ibm.com}
(2023)
\end{botherref}
\endbibitem

\bibitem{lucas2014ising}
\begin{barticle}
\bauthor{\bsnm{Lucas}, \binits{A.}}:
\batitle{Ising formulations of many np problems}.
\bjtitle{Frontiers in physics}
\bvolume{2},
\bfpage{5}
(\byear{2014})
\end{barticle}
\endbibitem

\end{thebibliography}

\newpage
\begin{appendices}

\section{Formulating the TSP as a QUBO}
\label{app:qubo}

In this section, we describe here how the TSP is encoded as a QUBO \cite{lucas2014ising}. A path $\sigma(1), \dotsc, \sigma(n)$ through the nodes $1, \dotsc, n$ of the graph is encoded by a set $x$ of $n^2$ binary variables $x_{i,p} \in \{ 0, 1\}$, where $i$ and $p$ are integers that range from $1$ to $n$. Let $x_{i,p} \in \{0,1\}$ be $1$ if the path goes through node $i$ at step $p$. The path visits $n$ nodes if and only if $\forall p \in \{ 1, \dotsc, n\}: \sum_{i=1}^n x_{i,p} = 1$. The path visits each node once if and only if $\forall i \in \{ 1, \dotsc, n\}: \sum_{p=1}^n x_{i,p} = 1$. The total distance to be minimized is $D(x) = \sum_{i=1}^n \sum_{j=1}^n d_{i,j} \sum_{p=1}^{n-1} x_{i,p}x_{j,p+1}$. The TSP is then equivalent to minimizing
    $$ C(x) = \sum_{i=1}^n \sum_{j=1}^n d_{i,j} \sum_{p=1}^{n-1} x_{i,p}x_{j,p+1} + A \sum_{p=1}^n \left( 1 - \sum_{i=1}^n x_{i,p} \right)^2 + A \sum_{i=1}^n \left( 1 - \sum_{p=1}^n x_{i,p} \right)^2 $$ as long as the penalty $A$ is large enough ($A > \max\{ \lvert d_{i,j} \rvert \}$).

\section{Definition of puzzles}
\label{app:puzzles}

\begin{itemize}
    \item PUZZLE1 (6 pieces):
        \begin{itemize}
            \item $P_0$ = [(0, 0), (400, 0), (400, 400)]
            \item $P_1$ = [(0, 0), (450, 0), (480, 470), (0, 480), (0, 400), (300, 400), (400, 300), (300, 200), (0, 200)]
            \item $P_2$ = [(0, 0), (100, 0), (100, 400), (200, 400), (200, 500), (0, 500)]
            \item $P_3$ = [(0, 0), (400, 0), (400, 280), (20, 690)]
            \item $P_4$ = [(0, 0), (100, 0), (100, 470), (0, 490), (0, 280), (-300, 280), (-370, 200), (-300, 130), (0, 130)]
            \item $P_5$ = [(0, 0), (100, 0), (100, 400), (200, 400), (200, 500), (0, 500)]
        \end{itemize}
    \item PUZZLE2 (7 pieces):
        \begin{itemize}
            \item $P_0$  = [(0, 0), (200, 0), (200, 300), (-100, 300), (-100, 100), (0, 100)]
            \item $P_1$ = [(0, 0), (200, 0), (190, 150), (100, 100), (0, 150)]
            \item $P_2$ = [(0, 0), (300, 0), (300, 100), (200, 100)]
            \item $P_3$ = [(0, 0), (300, 0), (300, 190), (200, 190), (200, 100), (0, 100)]
            \item $P_4$ = [(0, 0), (150, 0), (200, 100), (150, 200), (0, 200), (-50, 150), (0, 100), (-50, 50)]
            \item $P_5$ = [(0, 0), (200, 0), (200, 90), (150, 40), (100, 90), (50, 40), (0, 90)]
            \item $P_6$ = [(0, 0), (300, 0), (300, 100), (200, 100)]
        \end{itemize}
    \item PUZZLE3 (12 pieces):
        \begin{itemize}
            \item $P_0$ = [(0, 0), (400, 0), (400, 120), (480, 120), (480, 480), (280, 480), (280, 720), (480, 720), (480, 780), (0, 780)]
            \item $P_1$ = [(0, 0), (700, 0), (700, 250), (600, 150), (500, 250), (300, 250), (200, 350), (100, 250), (0, 250)]
            \item $P_2$ = [(0, 0), (680, 0), (700, 380), (500, 180), (300, 380), (0, 170)]
            \item $P_3$ = [(0, 0), (270, 0), (170, 100), (270, 200), (170, 300), (170, 400), (0, 400), (0, 270), (-70, 270), (-70, 120), (0, 120)]
            \item $P_4$ = [(0, 0), (300, 0), (300, 500), (200, 370), (100, 650), (0, 470)]
            \item $P_5$ = [(0, 0), (280, 0), (180, 320), (280, 320), (280, 480), (180, 480), (280, 780), (0, 780)]
            \item $P_6$ = [(0, 0), (300, 0), (300, 580), (200, 280), (100, 700), (0, 500)]
            \item $P_7$ = [(0, 0), (100, 0), (100, 100), (0, 200), (100, 300), (100, 500), (200, 600), (100, 700), (-200, 700), (0, 400), (-200, 200)]
            \item $P_8$ = [(0, 0), (100, 0), (100, 80), (300, 80), (300, 0), (400, 0), (400, 200), (0, 200)]
            \item $P_9$ = [(0, 0), (400, 0), (500, 300), (400, 300), (400, 500), (500, 500), (400, 800), (100, 800), (100, 700), (-100, 700), (-100, 500), (100, 500), (100, 100), (0, 100)]
            \item $P_{10}$ = [(0, 0), (100, 0), (100, 100), (200, 100), (200, 0), (300, 0), (300, 100), (400, 100), (400, 300), (300, 300), (200, 200), (100, 300), (0, 200)]
            \item $P_{11}$ = [(0, 0), (180, 0), (180, 80), (70, 80), (70, 220), (180, 220), (180, 280), (80, 280), (80, 400), (0, 400)]
        \end{itemize}
\end{itemize}

\section{Final placements}
\label{app:final_placements}

\begin{figure}[h]
\centering
\subfloat[Set of $6$ pieces for PUZZLE1]{\includegraphics[width=6cm]{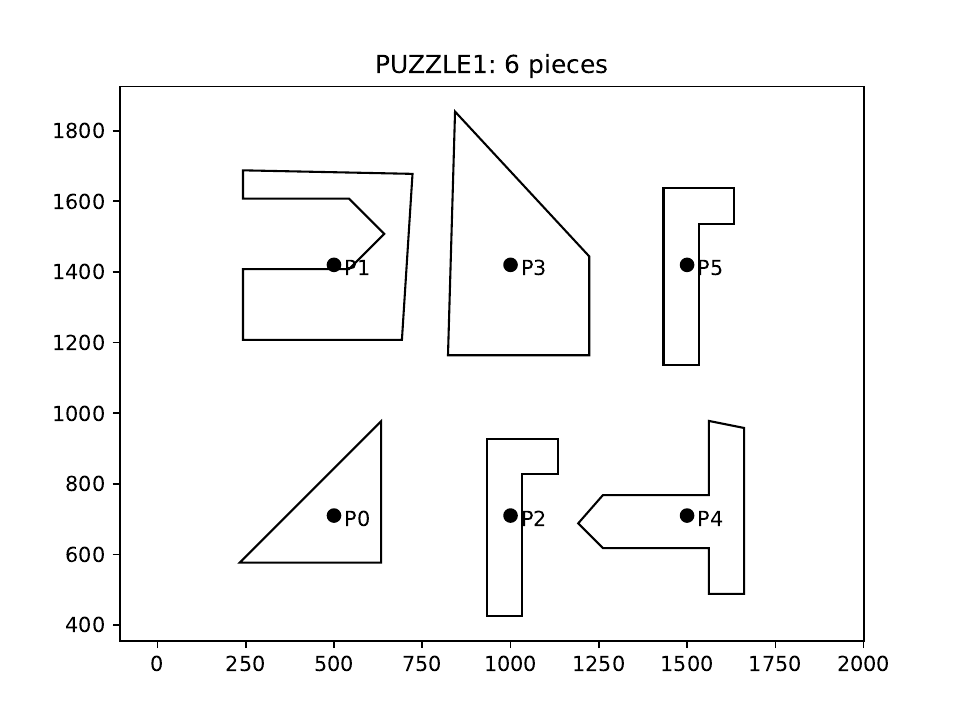}}
\subfloat[Opus Incertum (Brute Force Search)]{\includegraphics[width=6cm]{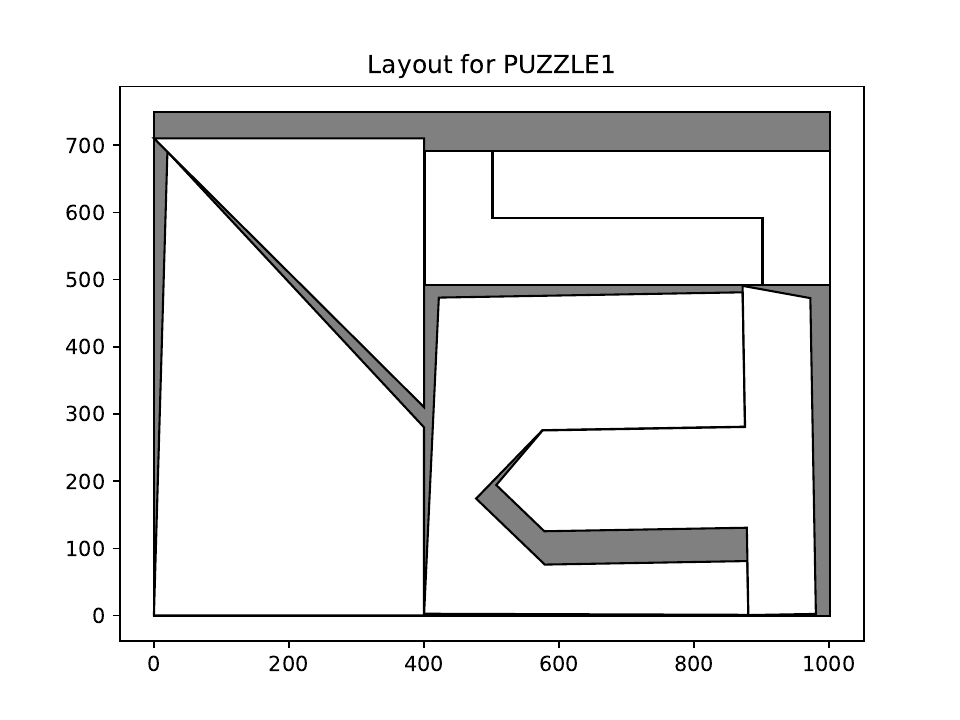}}

\subfloat[Set of $7$ pieces for PUZZLE2]{\includegraphics[width=6cm]{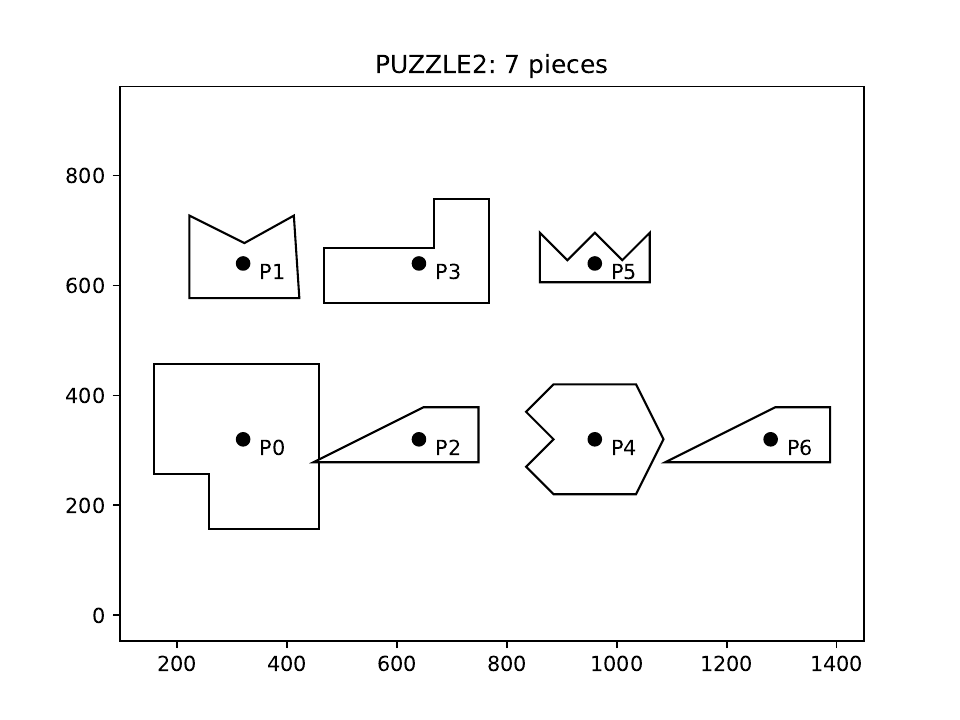}}
\subfloat[Opus Incertum (Brute Force Search)]{\includegraphics[width=6cm]{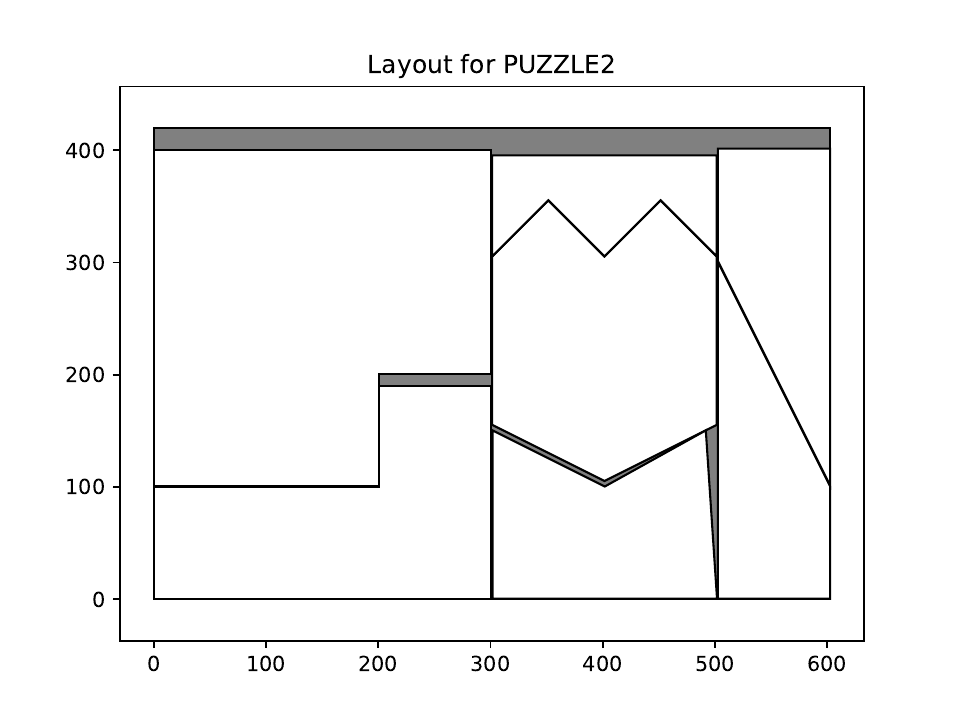}}

\subfloat[Set of $12$ pieces for PUZZLE3]{\includegraphics[width=6cm]{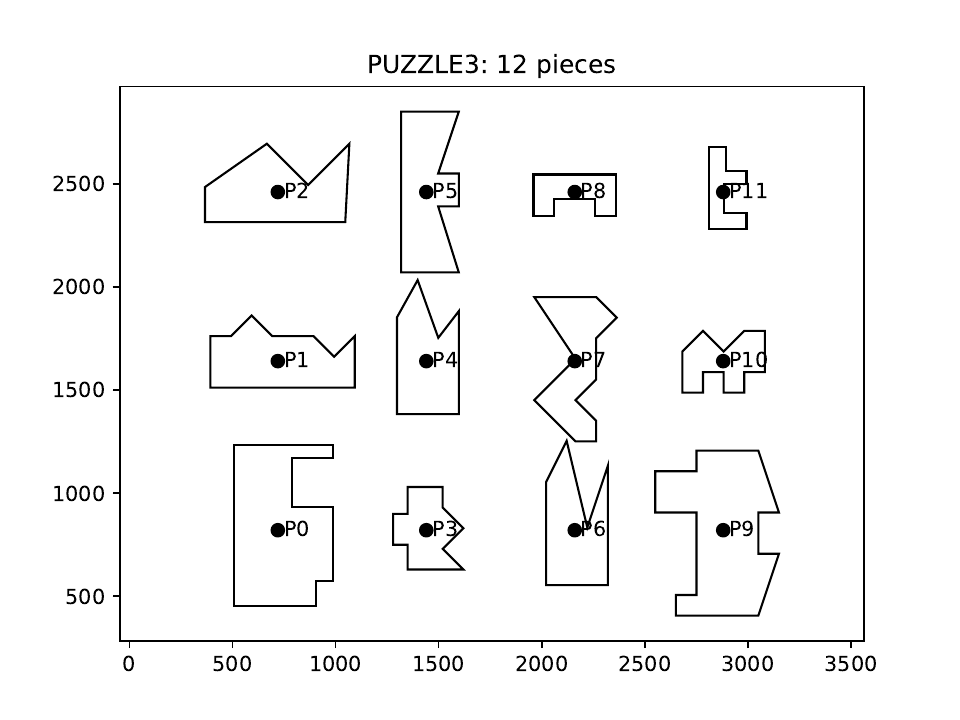}}
\subfloat[Opus Incertum (Brute Force Search)]{\includegraphics[width=6cm]{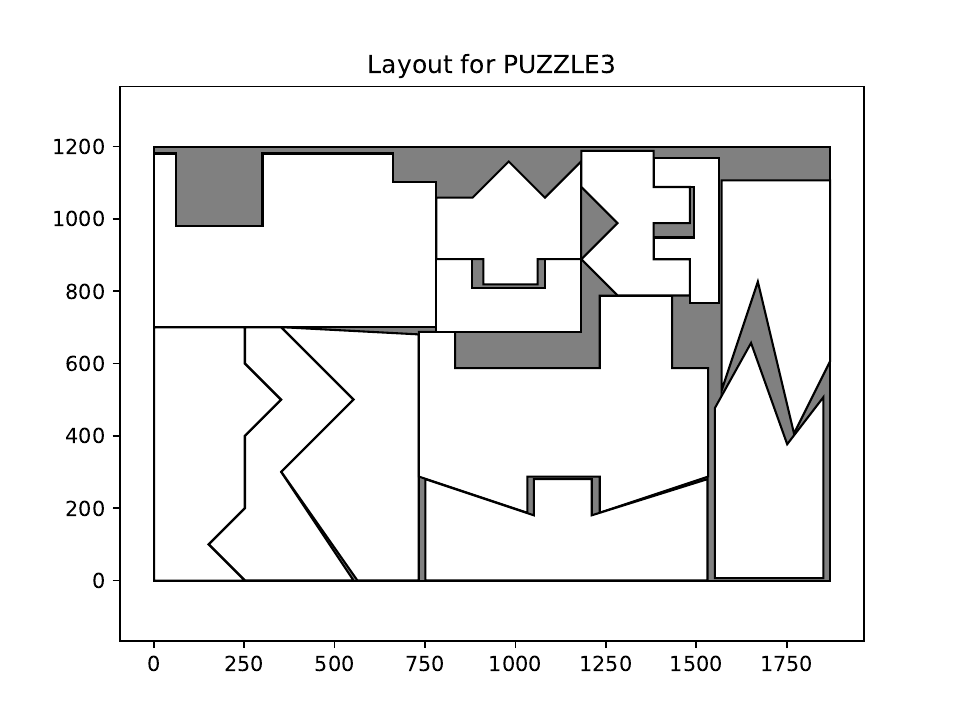}}
\caption{Results obtained for PUZZLE1, PUZZLE2 and PUZZLE3 using Opus Incertum and Brute Force Search.}
\label{fig:placements_brute_force}
\end{figure}

\begin{figure}[h]
\centering
\subfloat[Set of $43$ pieces for SHAPES1]{\includegraphics[width=6cm]{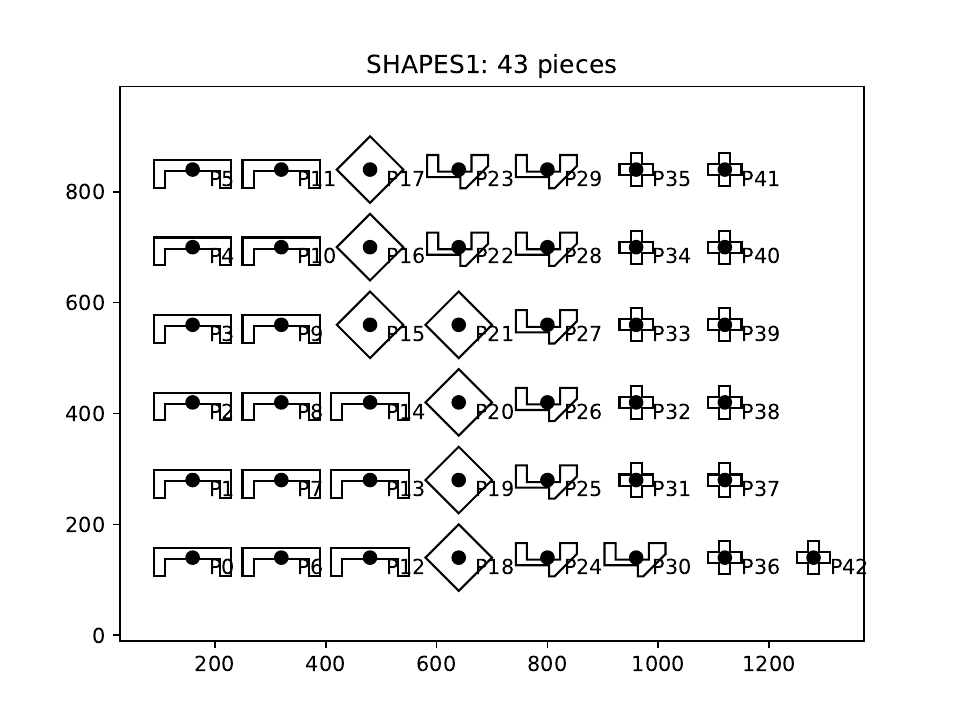}}
\subfloat[Opus Incertum (Brute Force Search)]{\includegraphics[width=6cm]{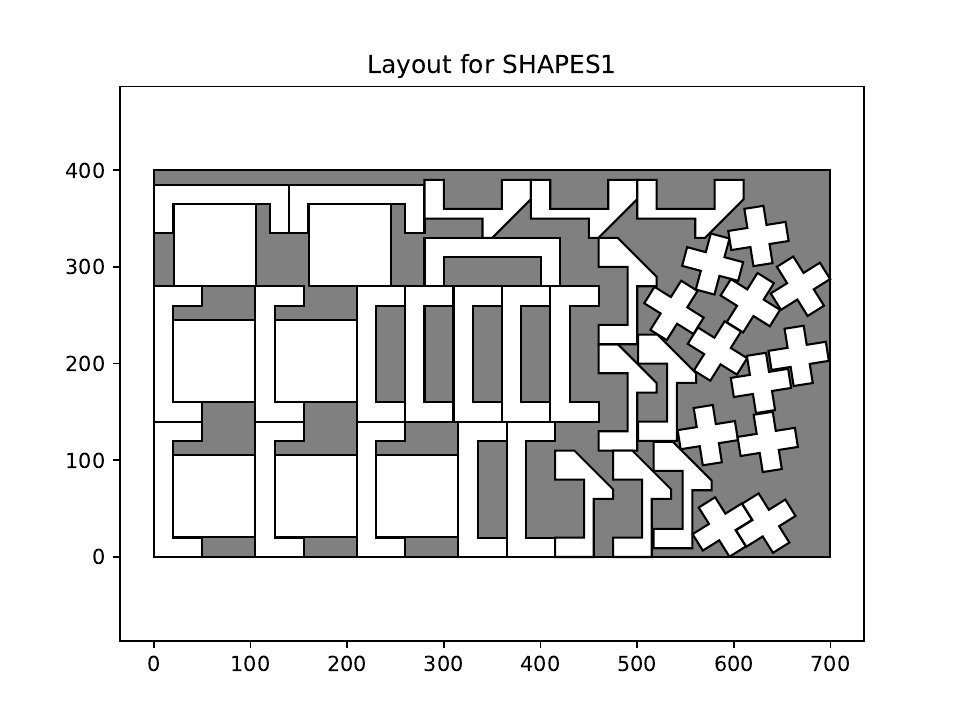}}

\subfloat[Opus Incertum (Quantum Approximate Optimization Algorithm, simulator)]{\includegraphics[width=6cm]{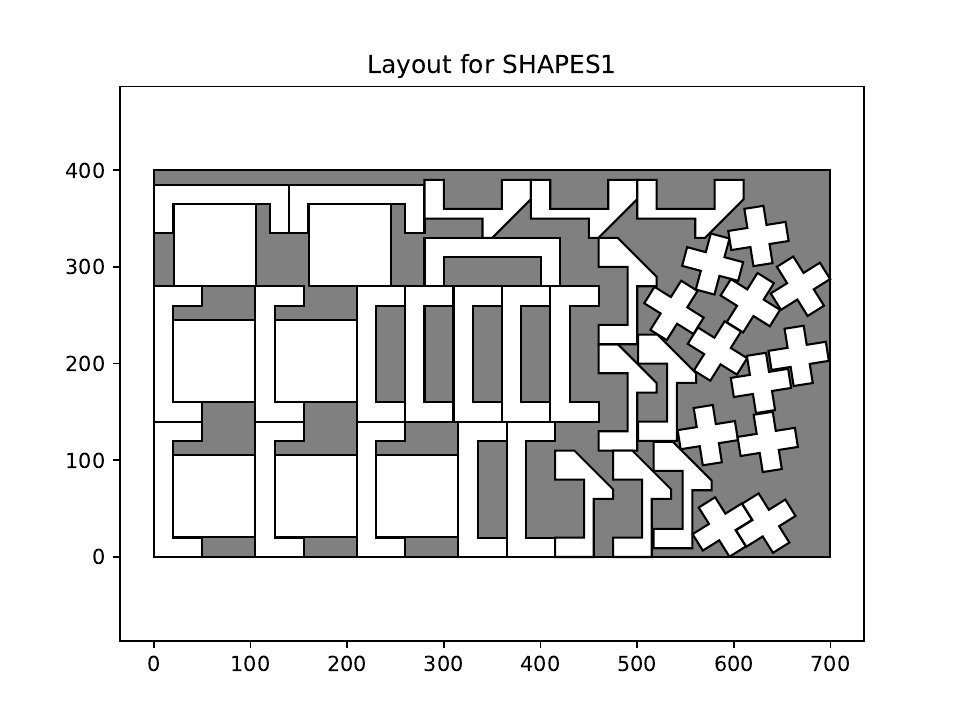}}
\subfloat[Opus Incertum (Quantum Approximate Optimization Algorithm, quantum computer)]{\includegraphics[width=6cm]{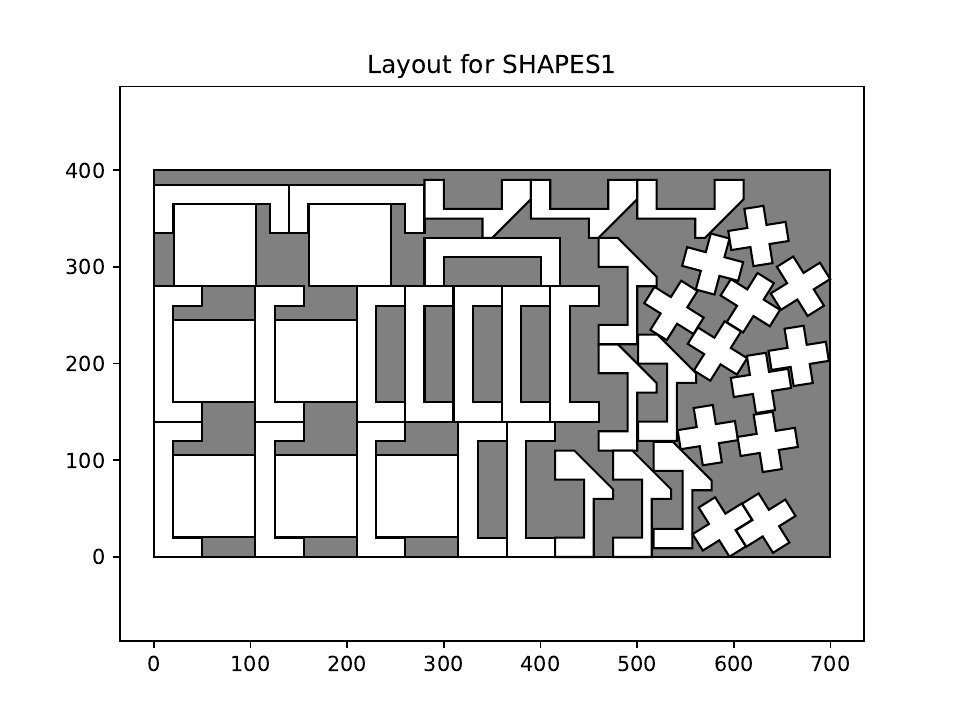}}

\subfloat[Opus Incertum (Quantum Alternating Operator Ansatz, simulator)]{\includegraphics[width=6cm]{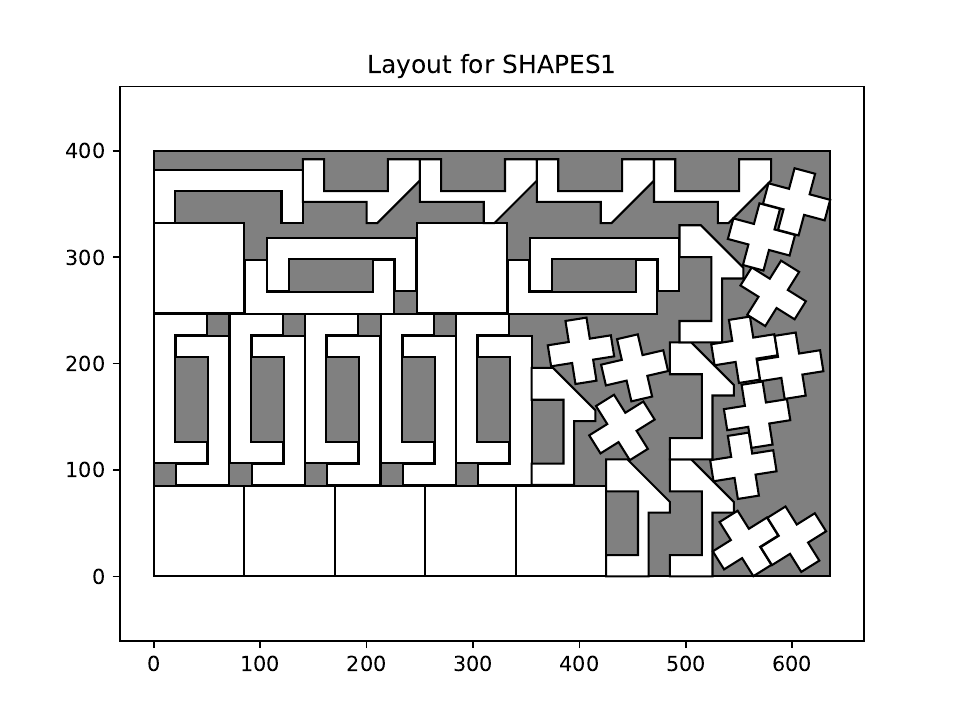}}
\subfloat[Opus Incertum (Quantum Alternating Operator Ansatz, quantum computer)]{\includegraphics[width=6cm]{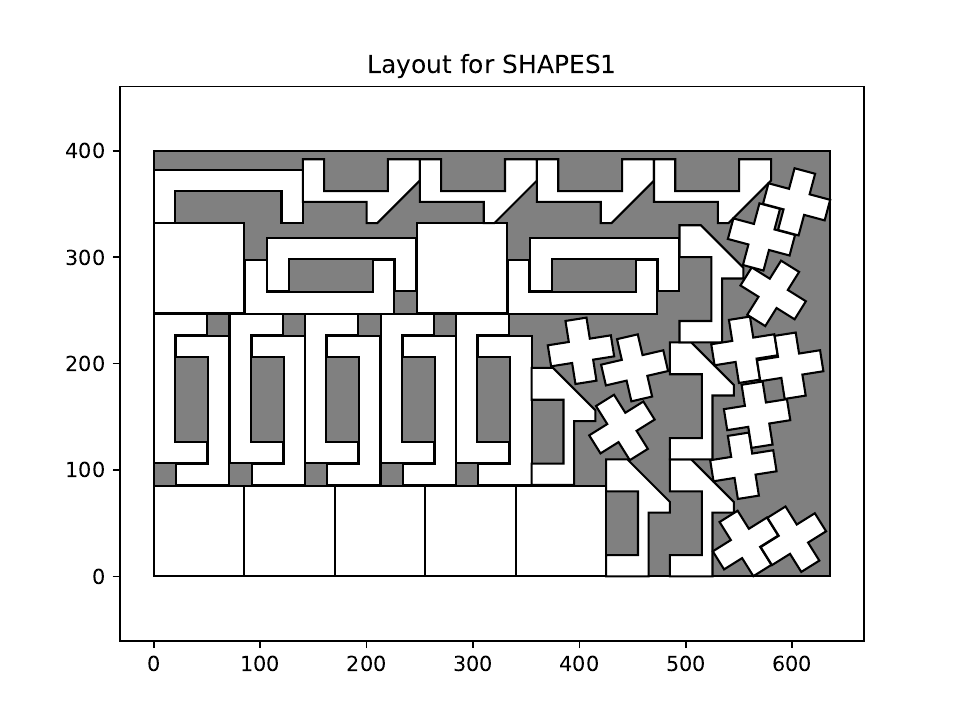}}
\caption{Results obtained for SHAPES1.}
\label{fig:placements_SHAPES1}
\end{figure}

\begin{figure}[h]
\centering
\subfloat[Set of $28$ pieces for SHAPES2]{\includegraphics[width=6cm]{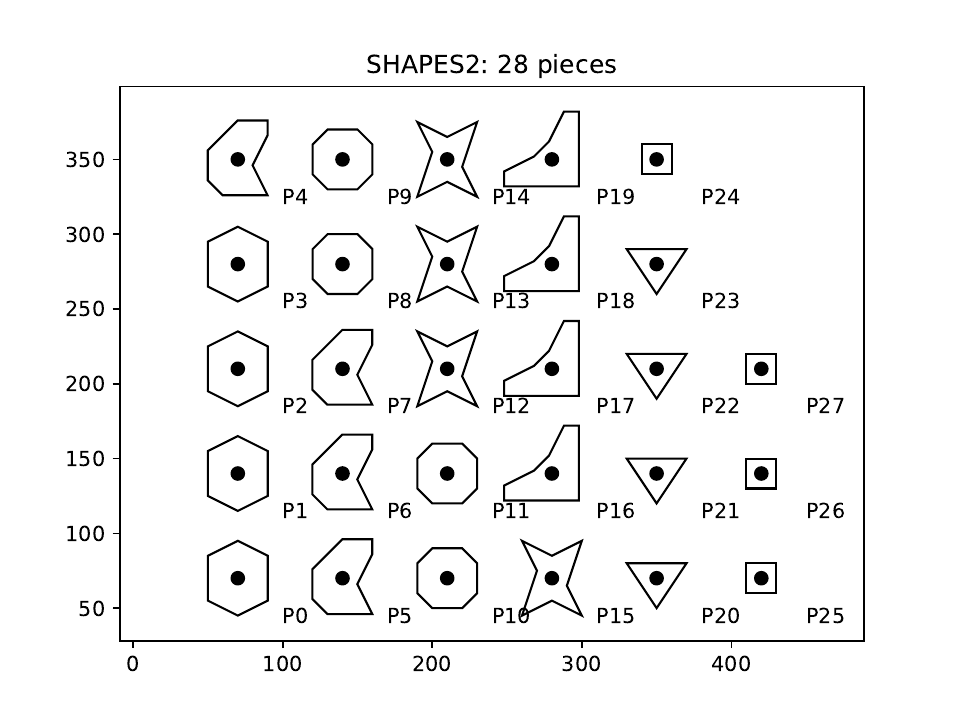}}
\subfloat[Opus Incertum (Brute Force Search)]{\includegraphics[width=6cm]{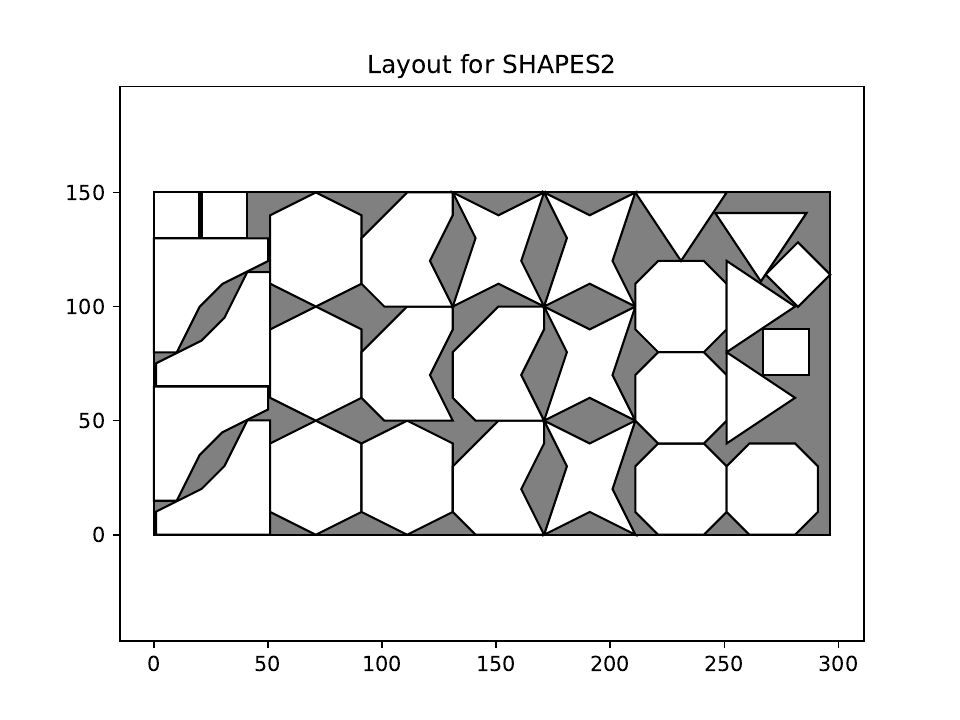}}

\subfloat[Opus Incertum (Quantum Approximate Optimization Algorithm, simulator)]{\includegraphics[width=6cm]{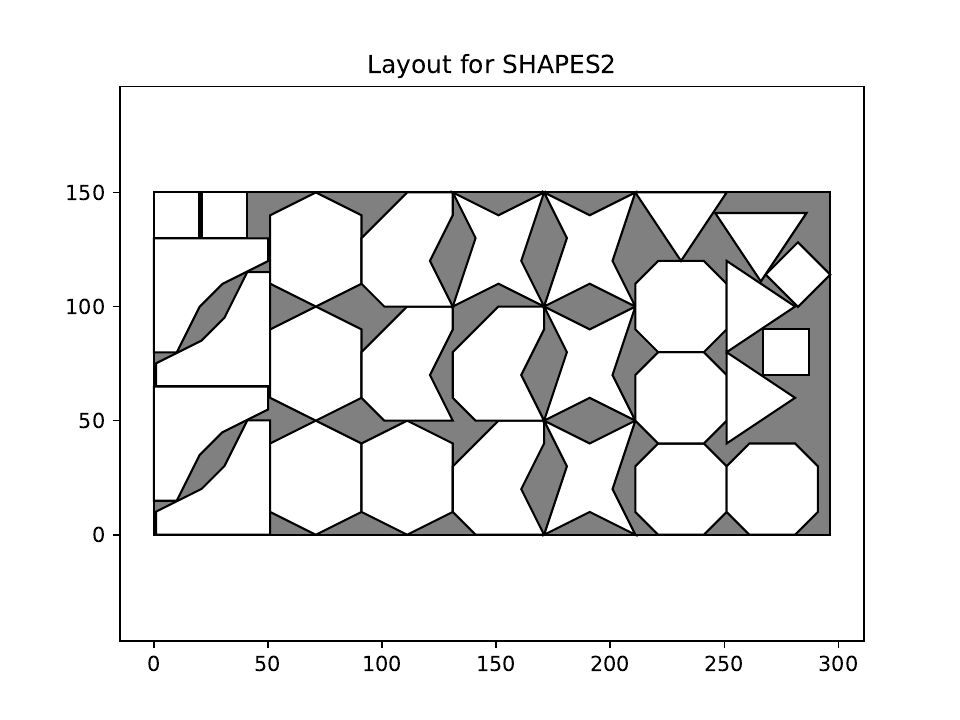}}
\subfloat[Opus Incertum (Quantum Approximate Optimization Algorithm, quantum computer)]{\includegraphics[width=6cm]{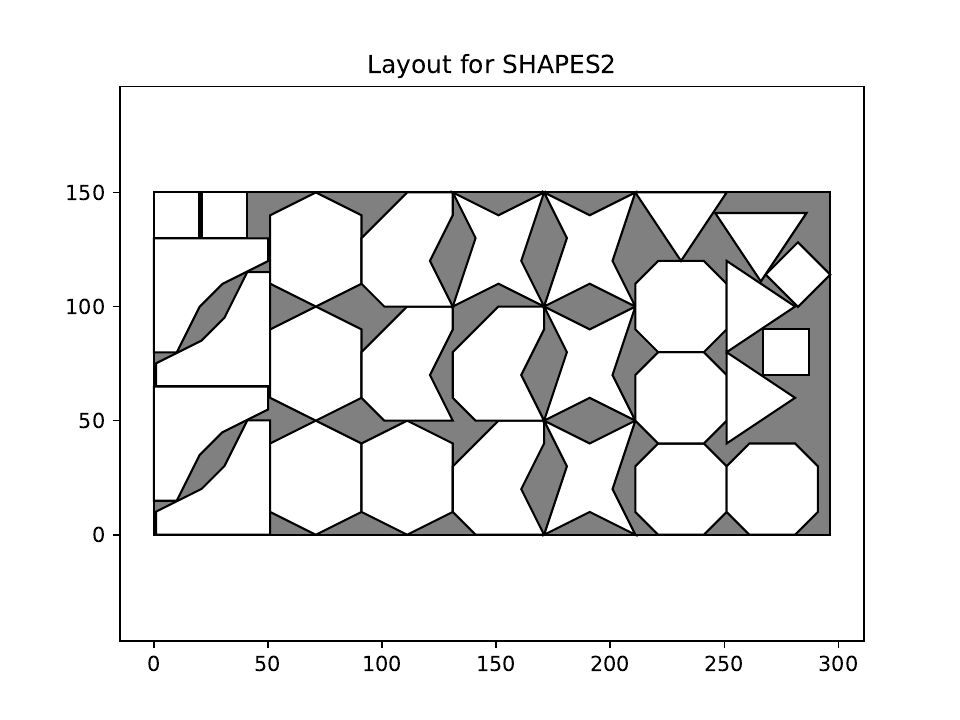}}

\subfloat[Opus Incertum (Quantum Alternating Operator Ansatz, simulator)]{\includegraphics[width=6cm]{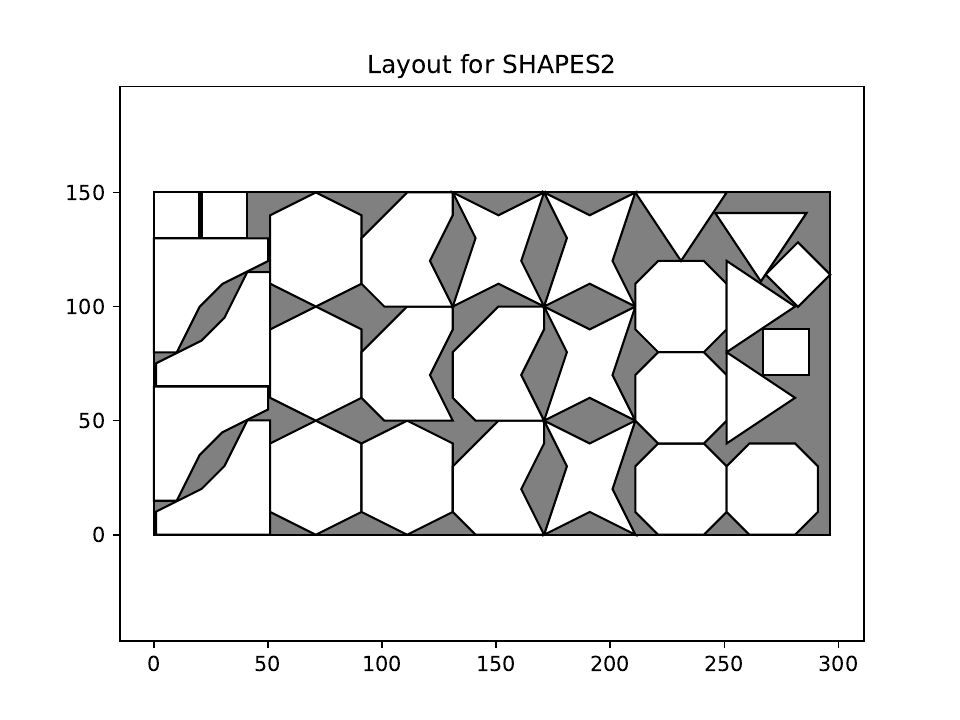}}
\subfloat[Opus Incertum (Quantum Alternating Operator Ansatz, quantum computer)]{\includegraphics[width=6cm]{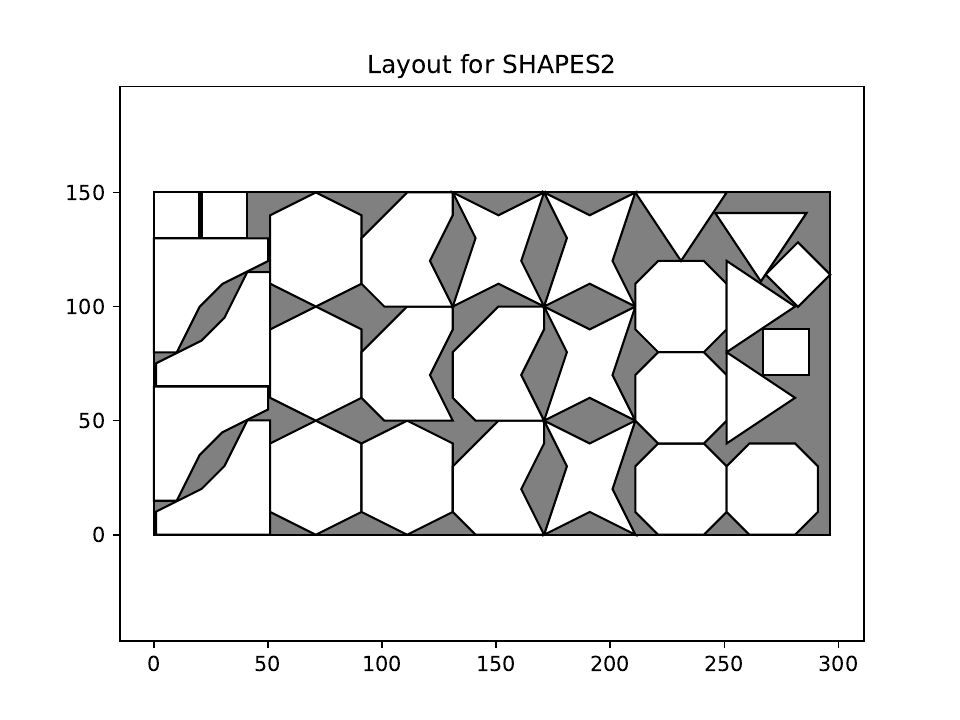}}
\caption{Results obtained for SHAPES2.}
\label{fig:placements_SHAPES2}
\end{figure}

\begin{figure}[h]
\centering
\subfloat[Set of $99$ pieces for SHIRTS]{\includegraphics[width=6cm]{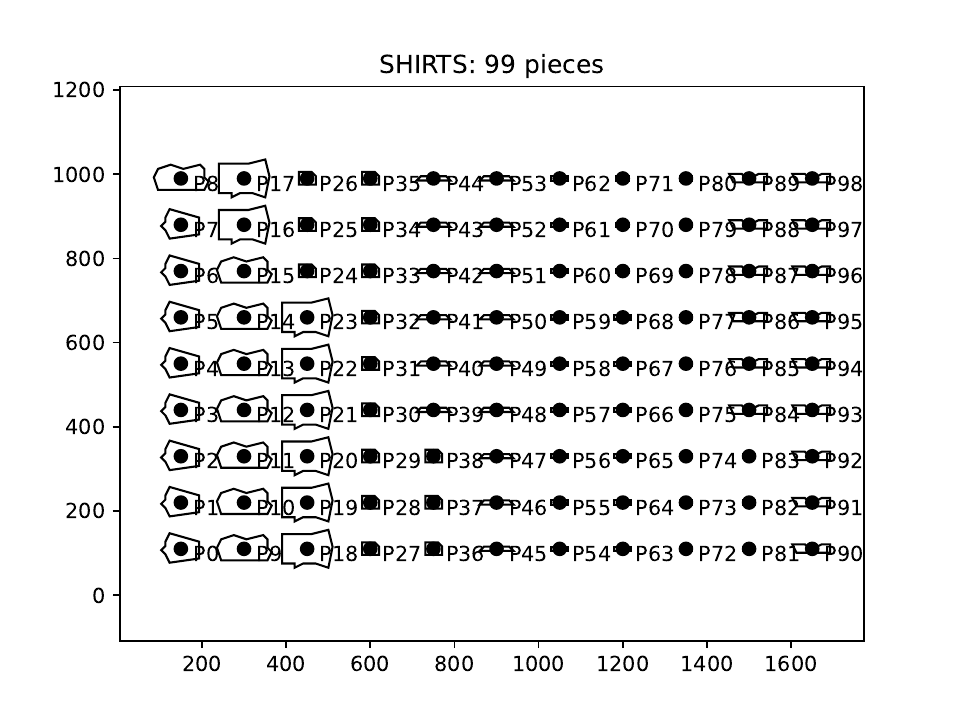}}
\subfloat[Opus Incertum (Brute Force Search)]{\includegraphics[width=6cm]{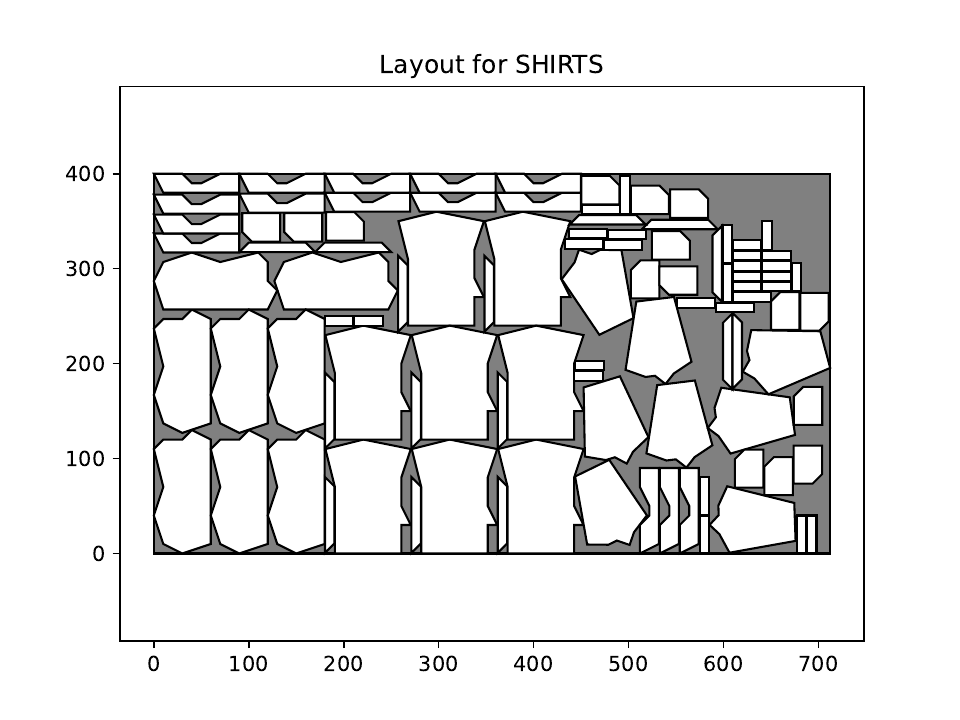}}

\subfloat[Opus Incertum (Quantum Approximate Optimization Algorithm, simulator)]{\includegraphics[width=6cm]{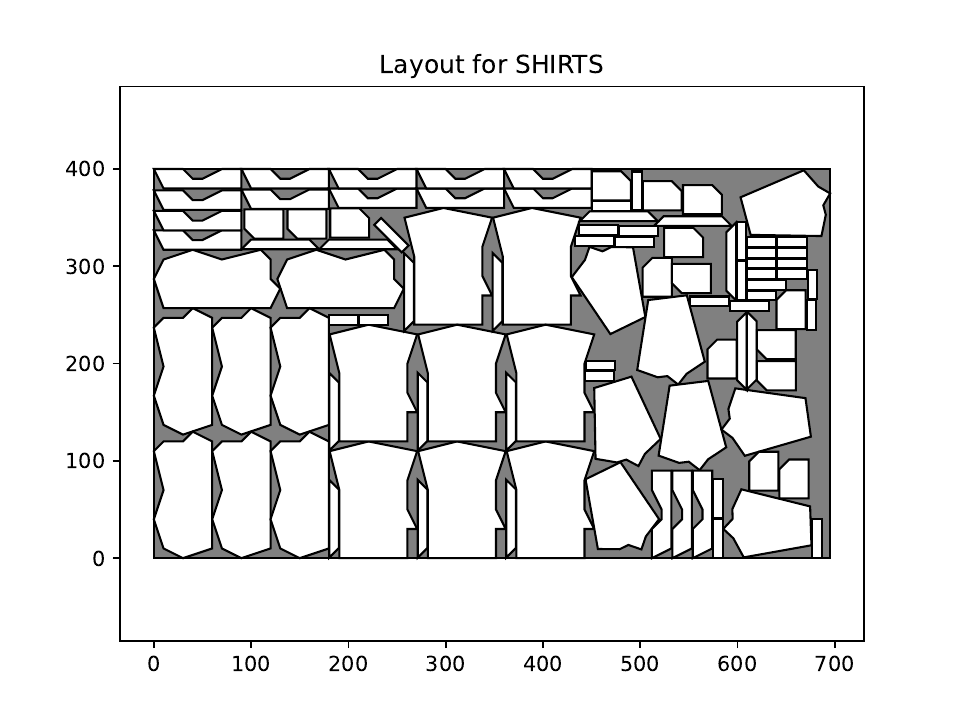}}
\subfloat[Opus Incertum (Quantum Approximate Optimization Algorithm, quantum computer)]{\includegraphics[width=6cm]{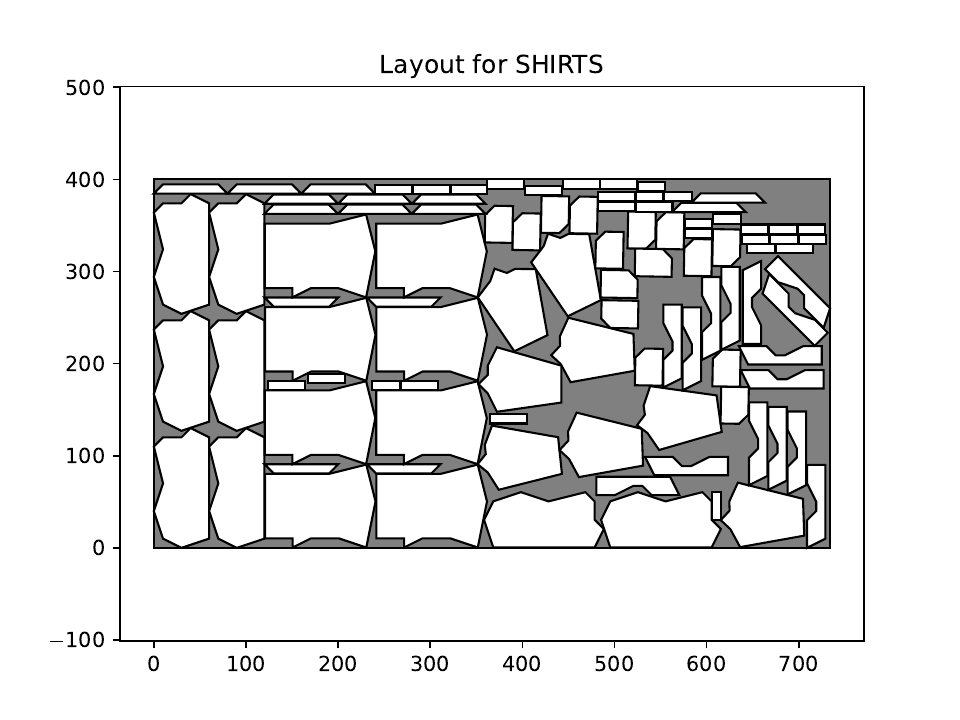}}

\subfloat[Opus Incertum (Quantum Alternating Operator Ansatz, simulator)]{\includegraphics[width=6cm]{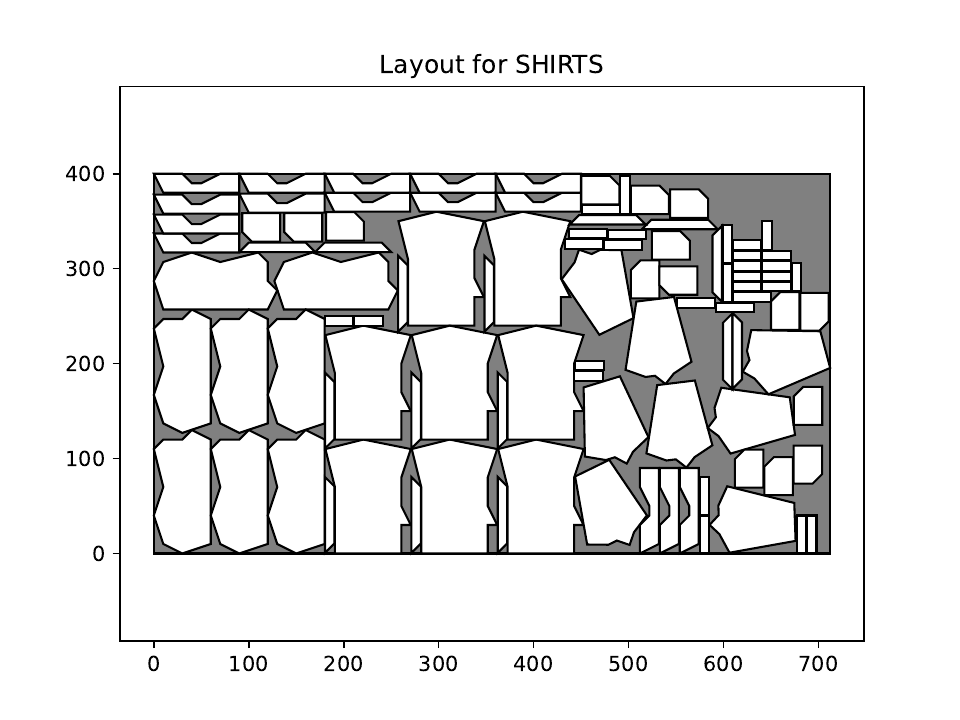}}
\subfloat[Opus Incertum (Quantum Alternating Operator Ansatz, quantum computer)]{\includegraphics[width=6cm]{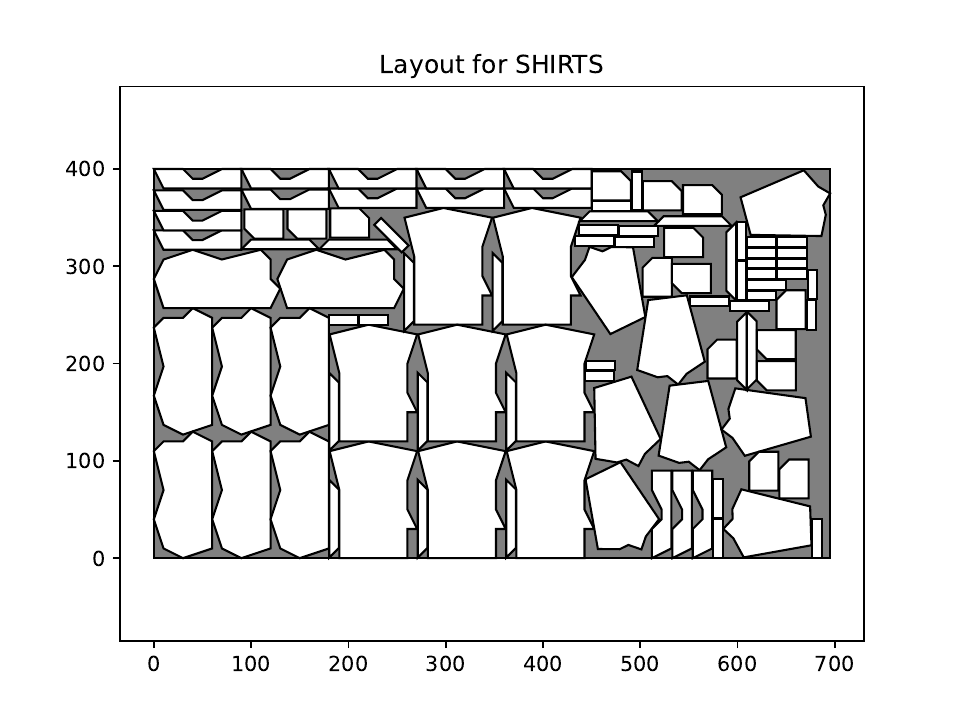}}
\caption{Results obtained for SHIRTS.}
\label{fig:placements_SHIRTS}
\end{figure}

\begin{figure}[h]
\centering
\subfloat[Set of $64$ pieces for SHIRTS]{\includegraphics[width=6cm]{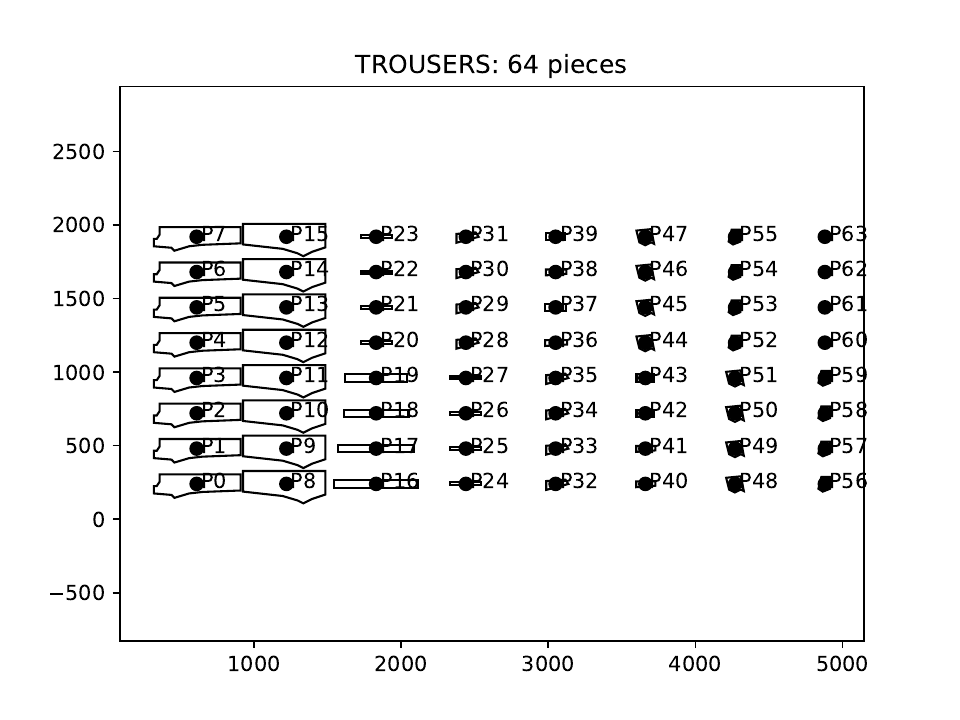}}
\subfloat[Opus Incertum (Brute Force Search)]{\includegraphics[width=6cm]{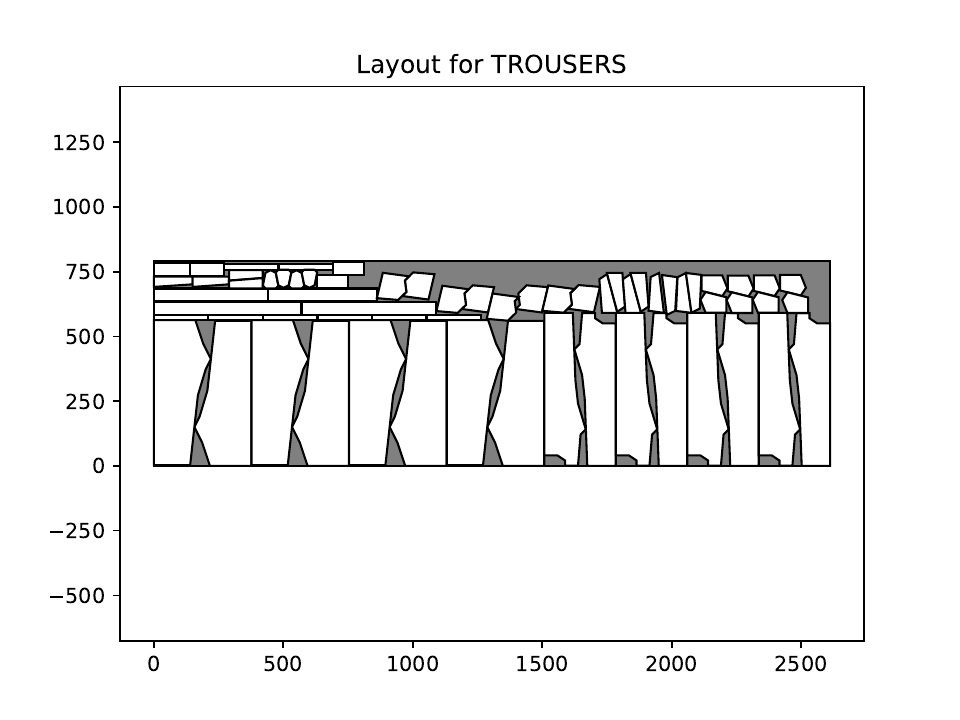}}

\subfloat[Opus Incertum (Quantum Approximate Optimization Algorithm, simulator)]{\includegraphics[width=6cm]{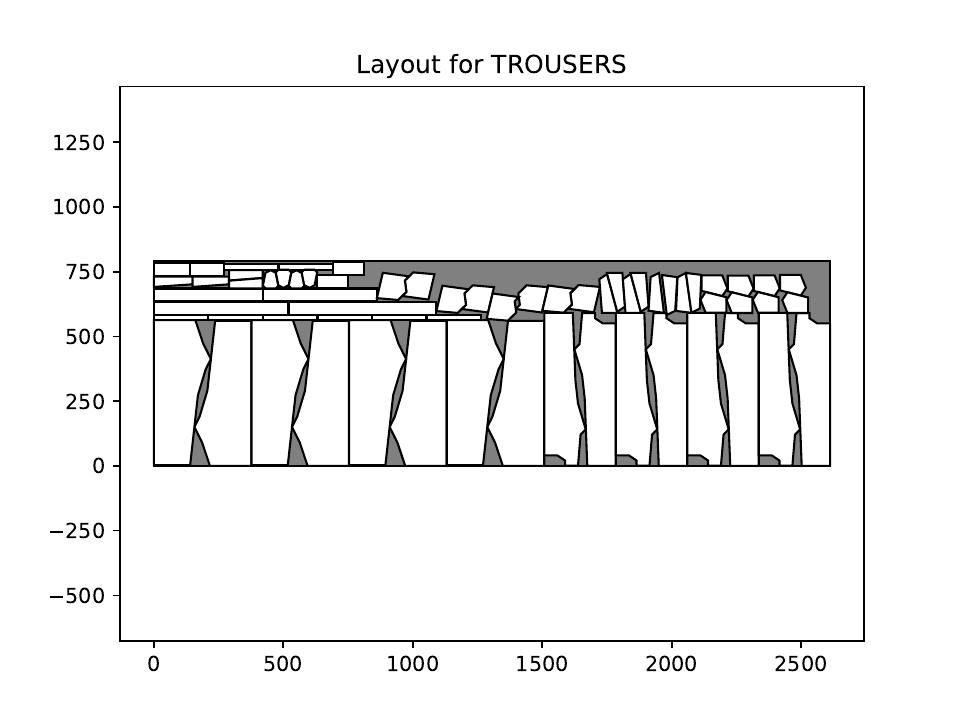}}
\subfloat[Opus Incertum (Quantum Approximate Optimization Algorithm, quantum computer)]{\includegraphics[width=6cm]{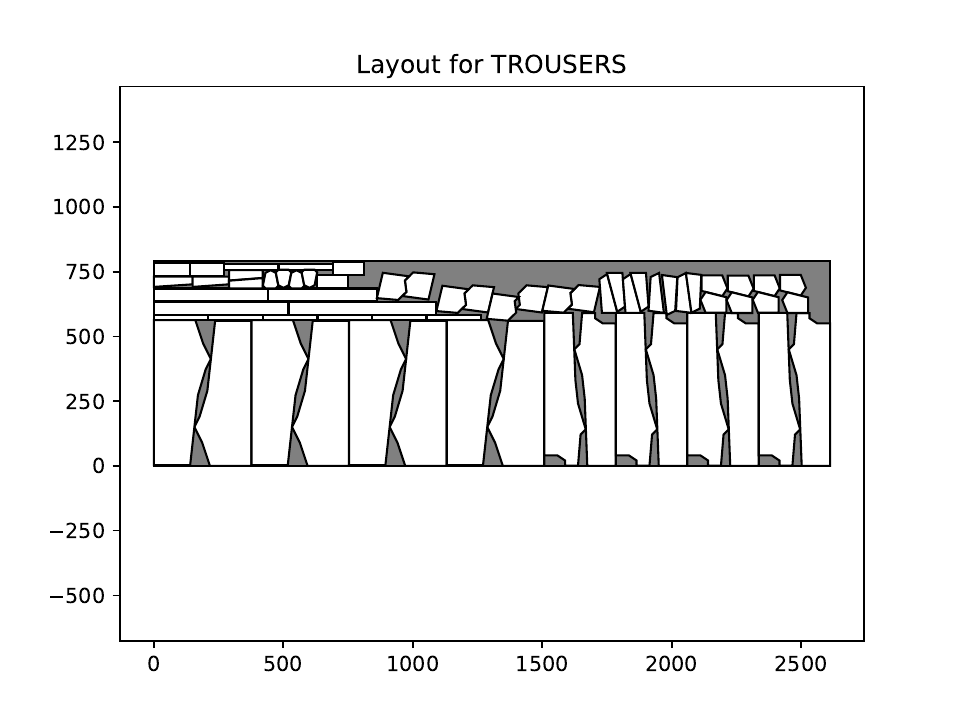}}

\subfloat[Opus Incertum (Quantum Alternating Operator Ansatz, simulator)]{\includegraphics[width=6cm]{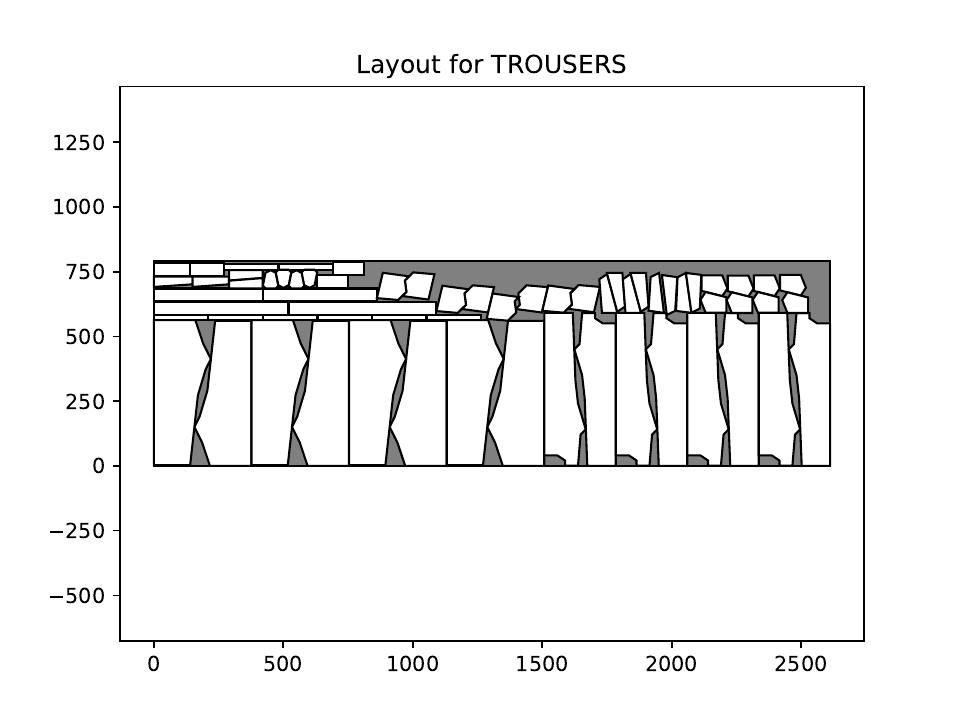}}
\subfloat[Opus Incertum (Quantum Alternating Operator Ansatz, quantum computer)]{\includegraphics[width=6cm]{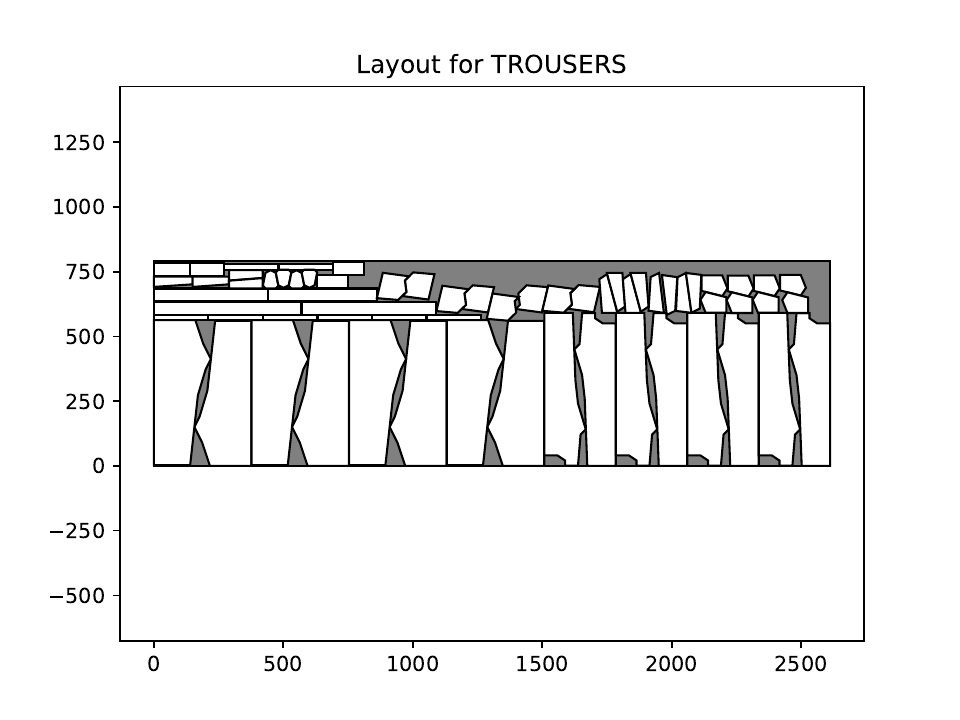}}
\caption{Results obtained for TROUSERS.}
\label{fig:placements_TROUSERS}
\end{figure}

\begin{figure}[h]
\centering
\subfloat[Set of $48$ pieces for SWIM]{\includegraphics[width=6cm]{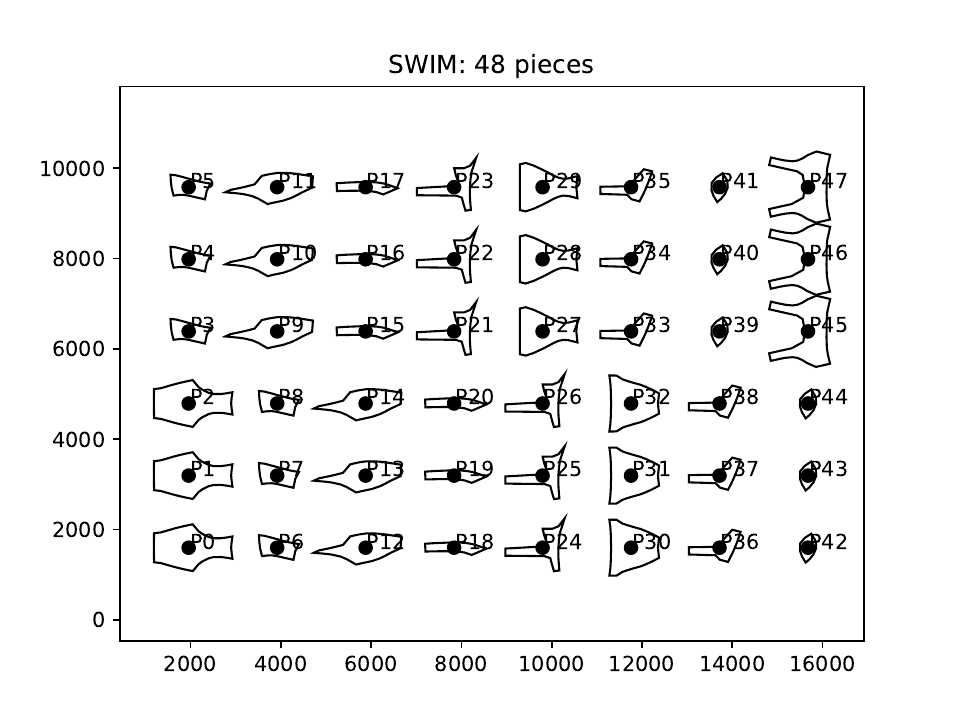}}
\subfloat[Opus Incertum (Brute Force Search)]{\includegraphics[width=6cm]{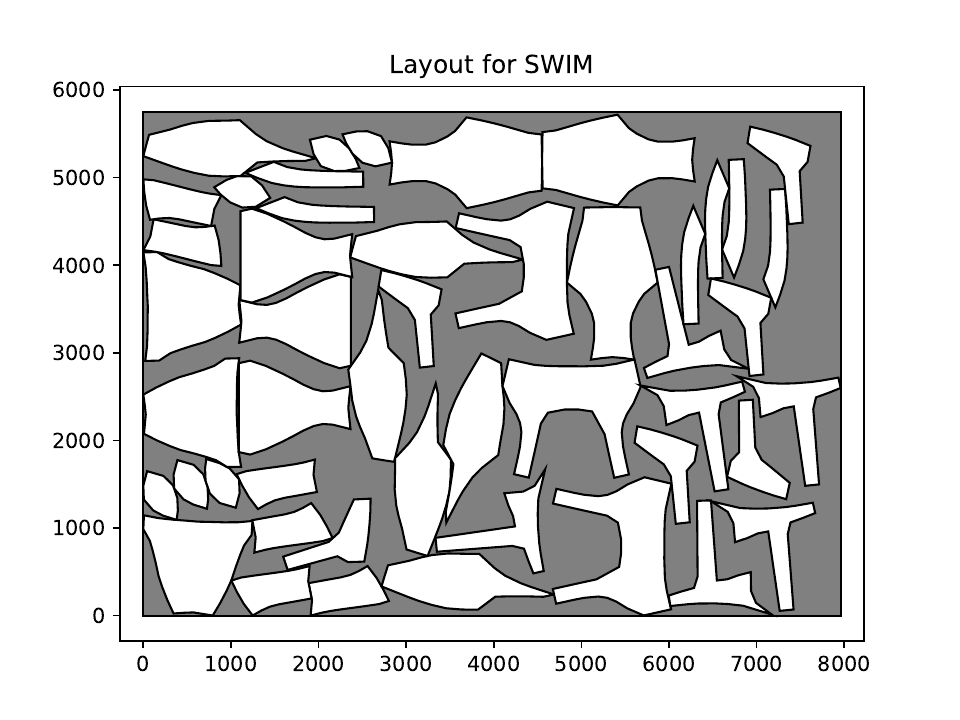}}

\subfloat[Opus Incertum (Quantum Approximate Optimization Algorithm, simulator)]{\includegraphics[width=6cm]{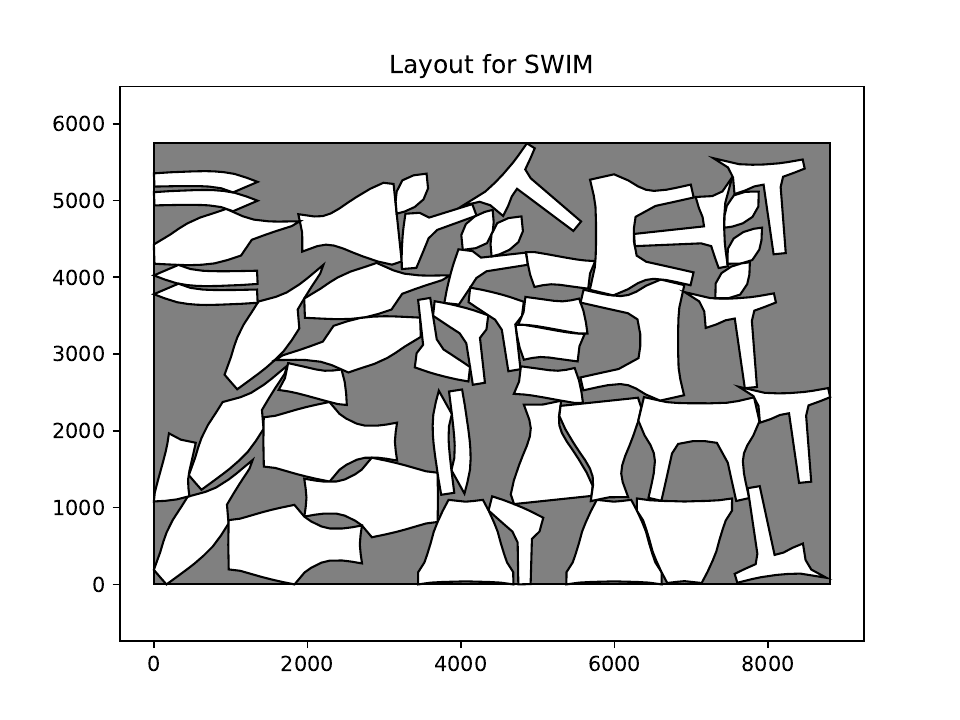}}
\subfloat[Opus Incertum (Quantum Approximate Optimization Algorithm, quantum computer)]{\includegraphics[width=6cm]{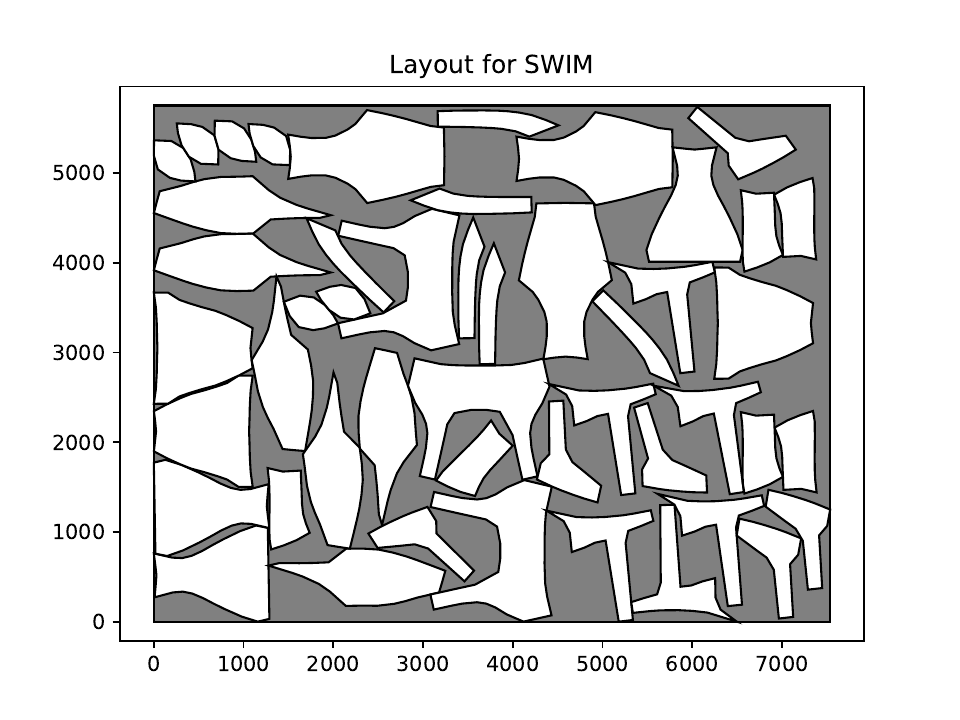}}

\subfloat[Opus Incertum (Quantum Alternating Operator Ansatz, simulator)]{\includegraphics[width=6cm]{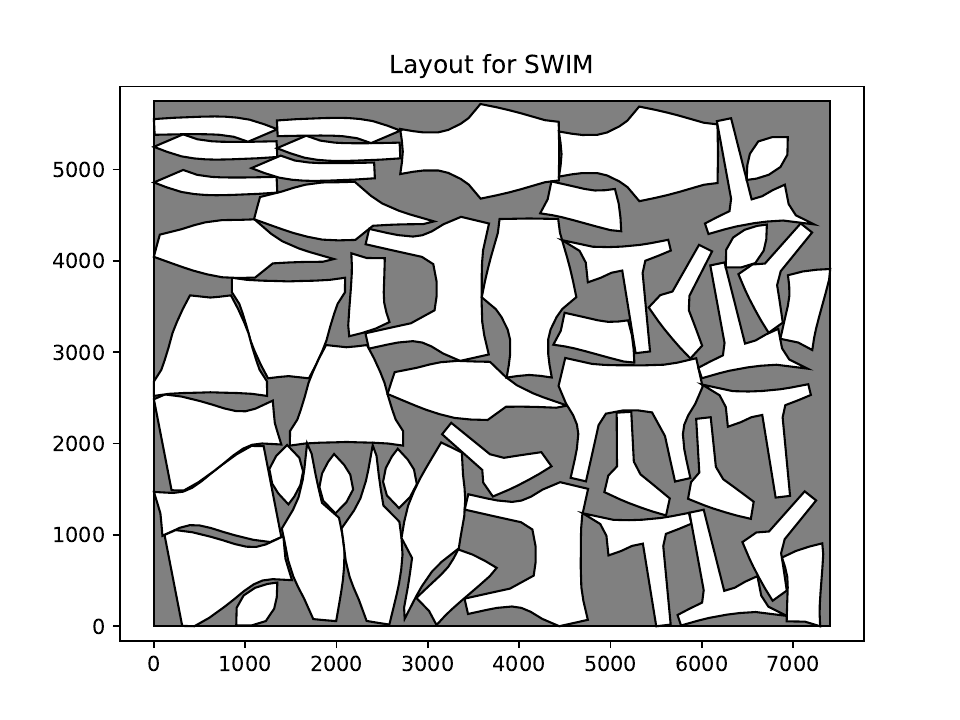}}
\subfloat[Opus Incertum (Quantum Alternating Operator Ansatz, quantum computer)]{\includegraphics[width=6cm]{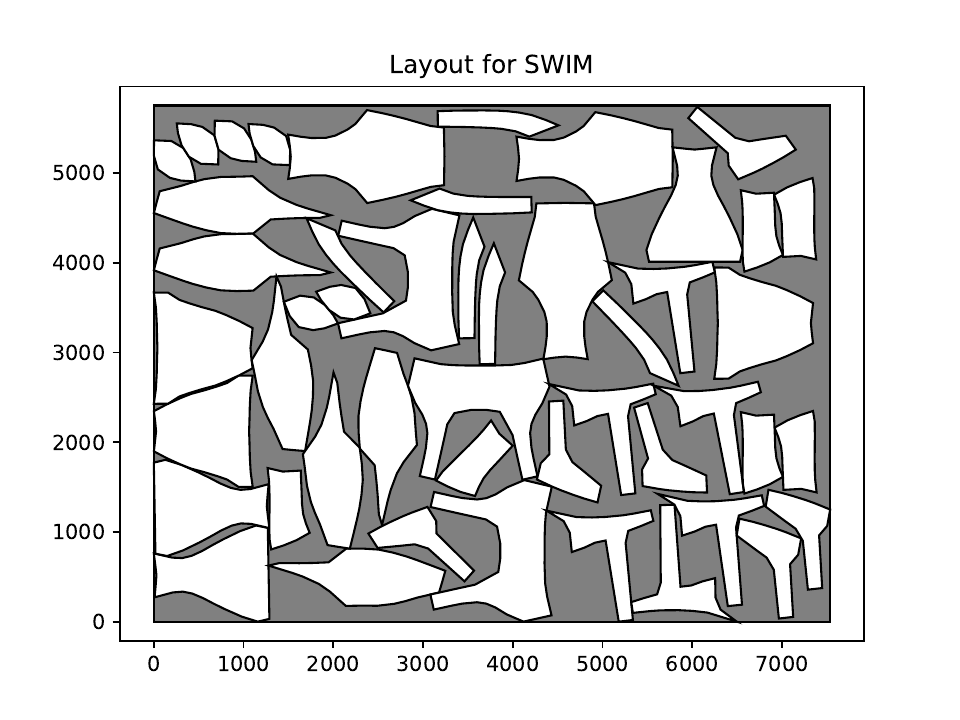}}
\caption{Results obtained for SWIM.}
\label{fig:placements_SWIM}
\end{figure}

\end{appendices}
\end{document}